\documentclass[showpacs,aps,amssymb,floatfix,prd,amsmath,preprintnumbers]{revtex4}
\usepackage{epstopdf}
\usepackage{capt-of}
\usepackage{ragged2e}
\usepackage{mathtools}
\usepackage{graphicx}  
\usepackage{dcolumn}   
\usepackage{bm}
\usepackage{hyperref}
\begin{document}
\input epsf.tex
\justify
\title{\bf $f(R,T)=f(R)+\lambda T$ gravity models as alternatives to cosmic acceleration}

\author{P.K. Sahoo$^{1}$\footnote{Email: pksahoo@hyderabad.bits-pilani.ac.in},  P.H.R.S. Moraes$^{2}$\footnote{Email: moraes.phrs@gmail.com}, Parbati Sahoo$^{1}$\footnote{Email: sahooparbati1990@gmail.com}, Binaya K. Bishi$^{3}$\footnote{Email: binaybc@gmail.com}}\

\affiliation{$^{1}$ Department of Mathematics, Birla Institute of
Technology and Science-Pilani, Hyderabad Campus, Hyderabad-500078,
India}
\affiliation{$^{2}$ ITA - Instituto Tecnol\'ogico de Aeron\'autica - Departamento de F\'isica, 12228-900, S\~ao Jos\'e dos Campos, S\~ao Paulo, Brasil}
\affiliation{$^{3}$ Department of Mathematics, Lovely Professional University, Phagwara, Jalandhar, Panjab-144401, India}

\begin{abstract}

This article presents cosmological models that arise in a subclass of $f(R,T)=f(R)+f(T)$ gravity models, with different $f(R)$ functions and fixed $T$-dependence. That is, the gravitational lagrangian is considered as $f(R,T)=f(R)+\lambda T$, with constant $\lambda$. Here $R$ and $T$ represent the Ricci scalar and trace of the stress-energy tensor, respectively. The modified gravitational field equations are obtained through the metric formalism for the Friedmann-Lema\^itre-Robertson-Walker  metric with signature $(+,-,-,-)$. We work with $f(R)=R+\alpha R^2-\frac{\mu^4}{R}$, $f(R)=R+k\ln(\gamma R)$ and $f(R)=R+me^{[-nR]}$, with $\alpha, \mu, k, \gamma, m$ and $n$ all free parameters, which lead to three different cosmological models for our Universe. For the choice of $\lambda=0$, this reduces to widely discussed $f(R)$ gravity models. This manuscript clearly describes the effects of adding the trace of the energy-momentum tensor in the $f(R)$ lagrangian. The exact solution of the modified field equations are obtained under the hybrid expansion law. Also we present the Om diagnostic analysis for the discussed models.
\end{abstract}

\keywords{$f(R,T)$ gravity; cosmic acceleration; hybrid expansion law; Om diagnostic}
\pacs{04.50.kd}

\maketitle

\section{Introduction}

The widely accepted theory of gravitation is the General Relativity (GR) theory, as it passed many experimental and observational tests. For example, recently, gravitational waves within the framework of GR were detected by LIGO and Virgo detectors \cite{Abbott/2016}.

Despite many attractive features including this great success, there are still several theoretical challenges, which motivate us to search for some modifications in GR. For example, GR does not provide us sufficient ideas to resolve some shortcomings like initial singularity, flatness issues, fine-tuning, cosmological constant and cosmic coincidence problems \cite{Sahni/2000,Carroll/2001,Peebles/2003,Padmanabhan/2003}. 

To overcome these problems, several modified theories are introduced in the literature. The importance of these theories for studying the behavior of the accelerating universe was investigated \cite{Nojiri/2007,Hu/2007,Appleby/2007,Starobinsky/2007}, in which modifications were made in the gravitational part of Einstein-Hilbert action. On the other hand, the matter part modification of Einstein-Hilbert action yields dynamical models such as quintessence, k-essence, Chaplygin gas and holographic dark energy models \cite{Zlatev/1999,Turner/2002,Sahni/2002,Chiba/2000,Setare/2007,Bernardini/2008,Hsu/2004,Li/2004,Bamba/2012}. These modified models can indeed well address the current accelerated expansion of the universe discovered by various observational aspects \cite{Bennett/2003,Perlmutter/1997,Riess/2007,Cole/2005,Eisenstein/2005,Jain/2003}.

One of the simplest modified theory is the $f(R)$ gravity, which is considered as most suitable for constructing cosmological models with differently ordered curvature invariants as a function of the Ricci scalar $R$. The unification of early-time inflation and late-time acceleration can be studied through $f(R)$ gravity models \cite{Nojiri/2008,Appleby/2010}. In the literature, it has been found that the higher order curvature terms in $f(R)$ gravity model play a vital role to avoid cosmological singularities \cite{Kanti/1999,Odintsov/2008,Bamba/2008}.

The weak field theory for stellar-like objects in the $f(R)$ theory of gravity was discussed in References \cite{Chiba/2007,Chiba/2008,Faraoni/2008}. Christian et al. in Reference \cite{Christian/2008} have shown that one can then find the behavior of $\psi(r)$ and $\phi(r)$ outside the star in the metric

\begin{widetext}
\begin{equation}
ds^{2}=-(1-2\psi(r))dt^{2}+(1-2\phi(r))dr^{2}+r^2(d\theta^{2}+\sin^2 \theta d\varphi^{2}),
\end{equation}
\end{widetext}
under the assumption that $f(R)$ is an analytic function at a constant curvature for a pressureless fluid, with $\psi(r)$ and $\phi(r)$ being the post Newtonian metric potentials. This analysis has led to the value of the post-Newtonian parameter $\gamma=\frac{1}{2}$, whereas from the solar system observations it is known that $\gamma=1$. This results contradicts most of the $f(R)$-type gravity models proposed in the literature so far. For example, models with $f(R)=R^{(1+\delta)}$, with $\delta\neq1$ conflict with the solar system test.

The $f(R)=R-\frac{\beta}{R^n}$-type models also suffer in passing the solar system tests \cite{Chiba/2003} and from gravitational instabilities \cite{Dolgov/2003}. Also these theories are incapable of producing standard matter dominated era followed by acceleration expansion \cite{Amendola/2007a,Amendola/2007b}. The $f(R)=R+\frac{\alpha}{R^m}-\frac{\beta}{R^n}$-type models have difficulties in satisfying the set of constraints coming from early and late-time acceleration, big bang nucleosynthesis and fifth-force experiments \cite{Brookfield/2006}. 

Due to all these issues corresponding to most of the $f(R)$ models, we will consider here some $f(R,T)$ models, for which $T$ is the trace of the energy-momentum tensor. Recently, the $f(R,T)$ gravity was developed by Harko et al. \cite{Harko/2011} as a generalization of $f(R)$ gravity. The theory contains an arbitrary function of the Ricci scalar $R$ along with the trace of energy-momentum tensor $T$.

Thereafter, a wide literature was developed in the context of $f(R,T)$ gravity, such as \cite{Houndjo/2012,Sharif/2012,Singh/2014,Moraes/2015,Zaregonbadi/2016,Sahoo/2016,Sahoo17,Shabani/2017,Sahoo/2017,Shabani017}. But, there are still so many cosmological questions to investigate in $f(R,T)$ gravity. In herein model we choose the following form for the $f(R,T)$ gravity function: $f(R,T)=f(R)+\lambda T$, with constant $\lambda$. That is, we fix the $T$-dependence of the theory on its simplest case while investigate different cases for the $R$-dependence of it. We shall investigate if the $T$-term is capable of evading the shortcomings one faces in $f(R)$ cosmological models. The accelerated expansion of the universe can indeed be described through modified gravity, but sometimes it faces a number of instabilities \cite{Chiba/2003,Dolgov/2003} which yields further modifications in cosmological models. 

Nojiri and Odinstov discussed a modified gravity with terms proportional to $\ln(R)$ or $R^{-n}(\ln R)^m$, which grow at small curvature \cite{Nojiri/2004}. The presence of $\ln(R)$ or $R^{-n}(\ln R)^m$ terms in $f(R)$ gravity may be responsible for the acceleration of the universe. Again, Nojiri and Odinstov discussed the $f(R)$ gravity cosmology by considering $f(R)=R+\gamma R^{-n}\left(\ln \frac{R}{\mu^2}\right)^m$ \cite{Nojiri/2007}. These forms for the $f(R)$ function were also used in \cite{Nojiri/2004,Nojiri/2007,Nozari/2009} to study different aspects of the theory. In \cite{Paul/2009}, the authors have shown that all these models exhibit current accelerating phase of the universe and the duration of the accelerating phase depends on the coupling constants of the gravitational action. 

Moreover, Yousaf et al. have explored the realistic configuration of anisotropic structure of compact stars in $f(R)$ gravity with three different forms for $f(R)$ \cite{Yousaf/2017}. 

In the present article we will consider three different choices for the $f(R)$ function as given in Ref.\cite{Paul/2009}. In the first model we will consider the mixed form for $f(R)$, namely a positive and a negative power of the curvature $R$, which is normally assumed to study the inflationary scenario of the early universe and the accelerating phase of the present universe. Such a functional form reads (A) $f(R,T)=R+\alpha R^2-\frac{\mu^4}{R}+\lambda T$, where the constants $\alpha$ and $\mu$ have dimension of $R^{-1}$ (i.e., $(time)^2$) and $R^{\frac{1}{2}}$ (i.e., $ (time)^{-1}$) \cite{Barrow/1983, Capozziello/2003}. The models (B) and (C) will be followed as $f(R,T)=R+k\ln(\gamma R)+\lambda T$ and $f(R,T)=R+me^{[-nR]}+\lambda T$ where $k, \gamma, m$ and $n$ are constants. Note that a form that allow a coupling between $R$ and $T$, such as $f(R,T)=R+\lambda T$, was already investigated in \cite{ms/2017}. 

\section{Basic Formalism of the $f(R,T)$ Gravity}

The modified Einstein-Hilbert action for the $f(R,T)$ gravity is given by \cite{Harko/2011}

\begin{equation}\label{e1}
S=\int \sqrt{-g}\biggl[\frac{1}{16\pi G}f(R,T)+L_{m}\biggr]d^{4}x,
\end{equation}
where $L_{m}$ is the usual matter Lagrangian density of matter, $f(R,T)$ is an arbitrary function of 
$R$ and $T$, the trace of the energy-momentum tensor $T_{ij}$ of matter, and $g$ is the determinant of the metric tensor $g_{ij}$. 

The energy-momentum tensor $T_{ij}$ from the Lagrangian matter is defined as

\begin{equation}\label{e2}
T_{ij}=g_{ij}L_{m}-\frac{\partial L_{m}}{\partial g^{ij}}.
\end{equation}

By varying action (\ref{e1}) with respect to the metric component, the $f(R,T)$ gravity field equations are obtained as

\begin{equation}\label{e3}
f_{R}(R,T)R_{ij}-\frac{1}{2}f(R,T)g_{ij}+(g_{ij}\Box -\nabla _{i}\nabla
_{j})f_{R}(R,T)=8\pi T_{ij}-f_{T}(R,T)T_{ij}-f_{T}(R,T)\Theta _{ij},
\end{equation}%
where
\begin{equation}\label{e4}
\Theta _{ij}=-2T_{ij}+g_{ij}L_{m}-2g^{lm}\frac{\partial ^{2}L_{m}}{\partial
g^{ij}\partial g^{lm}}.
\end{equation}
Here, $f_{R}(R,T)=\frac{\partial f(R,T)}{\partial R}$, $f_{T}(R,T)=\frac{%
\partial f(R,T)}{\partial T}$, $\Box \equiv \nabla ^{i}\nabla _{i}$, while $\nabla _{i}$ is the covariant derivative.

With the choice of $L_m=-p$, with $p$ being the pressure, and assuming units such that $G=1$, the term $\Theta_{ij}$ is given by $\Theta_{ij}=-2T_{ij}-pg_{ij}$ and Equation (\ref{e3}) reduces to
\begin{equation}\label{e5}
G_{ij}=T_{ij}^{eff}
\end{equation}
where
\begin{equation}\label{e6}
T_{ij}^{eff}=\frac{1}{f_{R}(R,T)}\left[(8\pi+f_{T}(R,T))T_{ij}+pf_{T}(R,T)g_{ij}+\frac{f(R,T)-Rf_R(R,T)}{2}g_{ij}-(g_{ij}\Box -\nabla _{i}\nabla
_{j})f_{R}(R,T)\right].
\end{equation}

\section{Field equations and Solutions}

In the present article, we will concentrate on a spatially flat Friedmann-Lema\^itre-Robertson-Walker universe with a time-dependent scale factor $a(t)$ such that the metric reads

\begin{equation}\label{e7}
ds^{2}=dt^{2}-a^{2}\left[dr^{2}+r^2(d\theta^{2}+\sin^2 \theta d\phi^{2})\right].
\end{equation}

The energy-momentum tensor for a perfect fluid, which will be assumed here, is written in the form
\begin{equation}\label{e8}
T_{ij}=(\rho+p)u_iu_j-pg_{ij},
\end{equation}
where $p$ and $\rho$ are, respectively, the pressure and energy density for the perfect fluid. Note that the trace of \eqref{e8} reads $T=\rho-3p$.

The general $f(R,T)$ gravity field equations for $f(R,T)=f(R)+\lambda T$ and the above metric is given by

\begin{equation}\label{e9}
3H^2=\frac{1}{f_{R}}\left[\left(8\pi +\frac{3\lambda}{2}\right)\rho-\frac{\lambda}{2} p\right]+\frac{1}{f_{R}}\left[\frac{f(R)-Rf_R}{2}-3H\dot{R}f_{RR}\right],
\end{equation}

\begin{equation}\label{e10}
2\dot{H}+3H^2=\frac{1}{f_{R}}\left[-\left(8\pi +\frac{3\lambda}{2}\right)p+\frac{\lambda}{2} \rho\right]-\frac{1}{f_{R}}\left[-\frac{f(R)-Rf_R}{2}+\dot{R}^2f_{RRR}+2H\dot{R}f_{RR}+\ddot{R}f_{RR}\right],
\end{equation}
with dots representing derivatives with respect to time $t$ and such that the Ricci scalar $R$ for metric (\ref{e7}) is

\begin{equation}\label{e11}
R=-6(\dot{H}+2H^2).
\end{equation}

From Equations (\ref{e9}) and (\ref{e10}), the pressure $p$, the energy density $\rho$ and the equation of state (EoS) parameter $\omega=p/\rho$ can be analytically expressed as

 \begin{multline}\label{e12}
\rho=\frac{f_R}{2}\left[\frac{-2\dot{H}}{8\pi+\lambda}+\frac{2\dot{H}+6H^2}{8\pi+2\lambda} \right]
+\left[\frac{H\dot{R}-\ddot{R}}{8\pi+\lambda}+\frac{5H\dot{R}+\ddot{R}}{8\pi+2\lambda}\right]\frac{f_{RR}}{2}+\left[\frac{\dot{R}^2}{8\pi+2\lambda}-\frac{\dot{R}^2}{8\pi+\lambda}\right]\frac{f_{RRR}}{2}-\frac{f(R)-Rf_R}{2(8\pi+2\lambda)},
\end{multline}

 \begin{multline}\label{e13}
p=\frac{f_R}{2}\left[\frac{-2\dot{H}}{8\pi+\lambda}-\frac{2\dot{H}+6H^2}{8\pi+2\lambda} \right]
+\left[\frac{H\dot{R}-\ddot{R}}{8\pi+\lambda}-\frac{5H\dot{R}+\ddot{R}}{8\pi+2\lambda}\right]\frac{f_{RR}}{2}+\left[\frac{-\dot{R}^2}{8\pi+2\lambda}-\frac{\dot{R}^2}{8\pi+\lambda}\right]\frac{f_{RRR}}{2}+\frac{f(R)-Rf_R}{2(8\pi+2\lambda)},
\end{multline}

\begin{equation}\label{e14}
\omega=\dfrac{\frac{f_R}{2}\left[\frac{-2\dot{H}}{8\pi+\lambda}-\frac{2\dot{H}+6H^2}{8\pi+2\lambda} \right]
+\left[\frac{H\dot{R}-\ddot{R}}{8\pi+\lambda}-\frac{5H\dot{R}+\ddot{R}}{8\pi+2\lambda}\right]\frac{f_{RR}}{2}+\left[\frac{-\dot{R}^2}{8\pi+2\lambda}-\frac{\dot{R}^2}{8\pi+\lambda}\right]\frac{f_{RRR}}{2}+\frac{f(R)-Rf_R}{2(8\pi+2\lambda)}}{\frac{f_R}{2}\left[\frac{-2\dot{H}}{8\pi+\lambda}+\frac{2\dot{H}+6H^2}{8\pi+2\lambda} \right]
+\left[\frac{H\dot{R}-\ddot{R}}{8\pi+\lambda}+\frac{5H\dot{R}+\ddot{R}}{8\pi+2\lambda}\right]\frac{f_{RR}}{2}+\left[\frac{\dot{R}^2}{8\pi+2\lambda}-\frac{\dot{R}^2}{8\pi+\lambda}\right]\frac{f_{RRR}}{2}-\frac{f(R)-Rf_R}{2(8\pi+2\lambda)}}.
\end{equation}

In order to derive exact solutions we will consider the hybrid expansion law for the scale factor as following \cite{Akarsu/2014}

\begin{equation}\label{e15}
a=t^{\eta}e^{\beta t},
\end{equation}
where $\eta$ and $\beta$ are positive constants. Such a scale factor yields the deceleration parameter and Hubble parameter as

\begin{equation}\label{e16}
q=-1+\frac{\eta}{(\beta t+\eta)^2},
\end{equation}
\begin{equation}\label{e16}
H=\frac{\eta +\beta  t}{t}.
\end{equation}

From the relation $a(t)=\frac{1}{1+z}$, with $z$ being the redshift and the present scale factor $a_0 = 1$, we obtain the following time-redshift relation:

\begin{equation}
t=\frac{\eta}{\beta} W\left[\beta\left(\frac{1}{z+1}\right)^{1/\eta}{\eta }\right],
\end{equation}
where $W$ denotes the Lambert function (also known as ``product logarithm'').

\begin{figure}[h!]
\minipage{0.50\textwidth}
  \includegraphics[width=75mm]{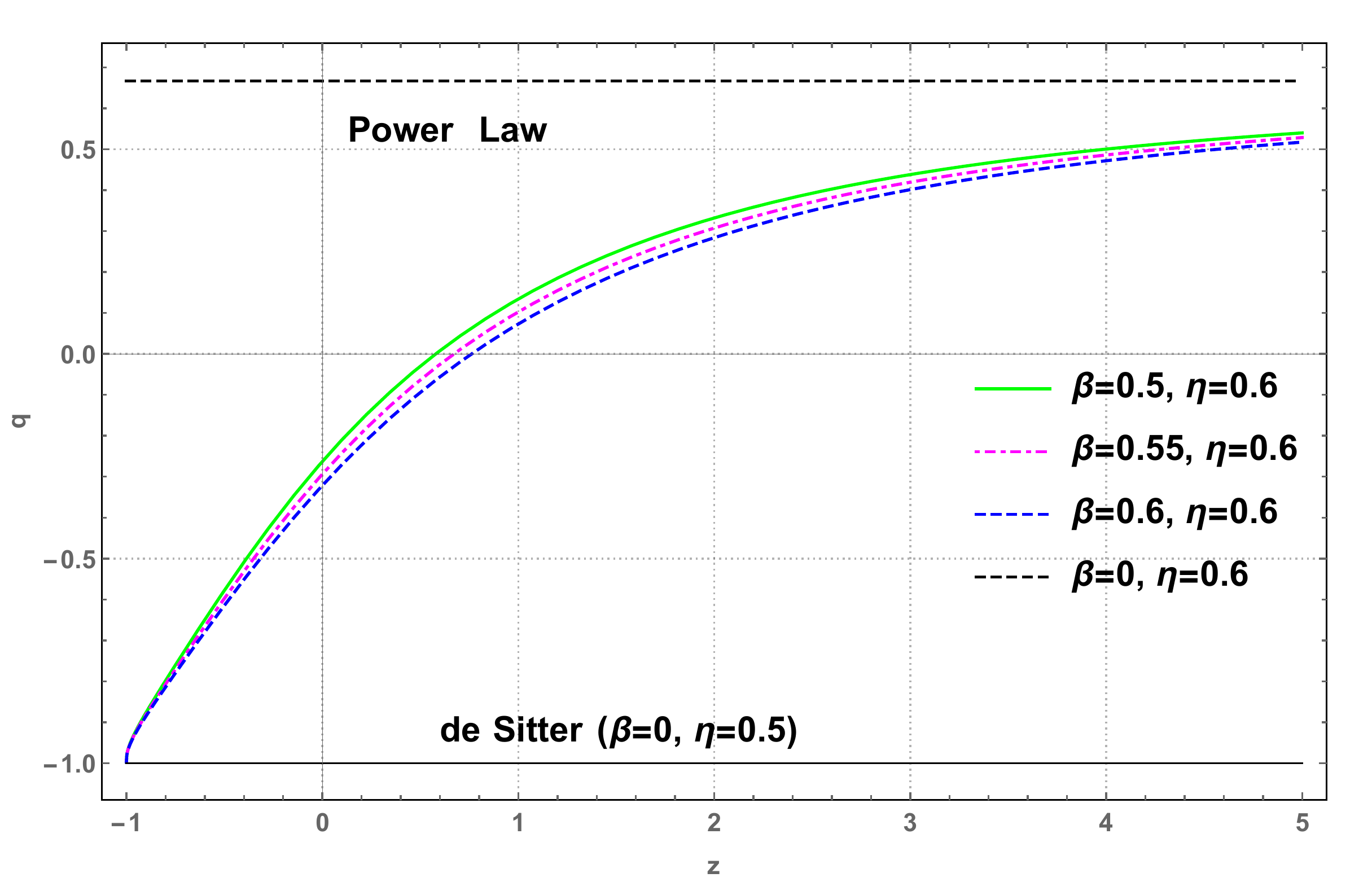}
   \caption{Variation of deceleration parameter $q$ against redshift $z$ .} \label{fig2}
\endminipage
\end{figure}

Plotting $q$ as a redshift function has the advantage of checking the
reliability of the model, through the redshift value in which the transition from the deceleration stage to the present acceleration era of the universe occurs. We will denote the transition redshift by $z_{tr}$. From Fig.1, the
transition occurs at $z_{tr} =0.5662,0.6691,0.7574$, corresponding to a fixed value for $\eta$, namely $\eta=0.6$, and various values for $\beta$, as $\beta=0.5,0.55,0.6$. The transition values for our model are in accordance with the observational data, as one can check in \cite{Capozziello14,Capozziello15,Farooq17}.

\subsection{The $f(R,T)=R+\alpha R^2-\frac{\mu^4}{R}+\lambda T$ Model}

In this case, by using Equation (\ref{e15}) for $f(R,T)=R+\alpha R^2-\frac{\mu^4}{R}+\lambda T$ in Equations (\ref{e12}-\ref{e14}), the analytical forms for $p$, $\rho$ and $\omega$ are expressed as follows:

\begin{multline}\label{e17}
\rho=\frac{1}{2t^2} \left[\frac{2 \eta }{\lambda +8 \pi }+\frac{3 (\eta +\beta  t)^2-\eta }{\lambda +4 \pi }\right] \left \{ \frac{\mu ^4 t^4}{36 \left[\eta -2 (\eta +\beta  t)^2\right]^2}-\frac{12 \alpha  \left[2 (\eta +\beta  t)^2-\eta \right]}{t^2}+1\right\}+\frac{3 \alpha G_{11}(t)}{t^4} \\+\frac{\mu ^4 t^2 G_{21}(t)}{36 \left[\eta -2 (\eta +\beta  t)^2\right]^4},
\end{multline}

\begin{multline}\label{e18}
p= \frac{1}{2t^2} \left[\frac{2 \eta }{\lambda +8 \pi }-\frac{3 (\eta +\beta  t)^2-\eta }{\lambda +4 \pi }\right] \left\{\frac{\mu ^4 t^4}{36 \left[\eta -2 (\eta +\beta  t)^2\right]^2}-\frac{12 \alpha  \left[2 (\eta +\beta  t)^2-\eta \right]}{t^2}+1\right\}\\
 -\frac{36 \alpha F_{11}(t) }{(\lambda +4\pi) (\lambda +8\pi) t^4} +\frac{\mu ^4 t^2 F_{21}(t) }{36 \left[\eta -2 (\eta +\beta  t)^2\right]^2},
\end{multline}

\begin{equation}\label{e19}
\omega=\dfrac{\frac{1}{2t^2} \left[\frac{2 \eta }{\lambda +8 \pi }-\frac{3 (\eta +\beta  t)^2-\eta }{\lambda +4 \pi }\right] \left\{\frac{\mu ^4 t^4}{36 \left[\eta -2 (\eta +\beta  t)^2\right]^2}-\frac{12 \alpha  \left(2 (\eta +\beta  t)^2-\eta \right)}{t^2}+1\right\}\\
 -\frac{36 \alpha F_{11}(t) }{(\lambda +4\pi) (\lambda +8\pi) t^4} +\frac{\mu ^4 t^2 F_{21}(t) }{36 \left[\eta -2 (\eta +\beta  t)^2\right]^2}}
{\frac{1}{2t^2} \left[\frac{2 \eta }{\lambda +8 \pi }+\frac{3 (\eta +\beta  t)^2-\eta }{\lambda +4 \pi }\right] \left \{ \frac{\mu ^4 t^4}{36 \left[\eta -2 (\eta +\beta  t)^2\right]^2}-\frac{12 \alpha  \left[2 (\eta +\beta  t)^2-\eta \right]}{t^2}+1\right\}+\frac{3 \alpha G_{11}(t)}{t^4} \\+\frac{\mu ^4 t^2 G_{21}(t)}{36 \left[\eta -2 (\eta +\beta  t)^2\right]^4}},
\end{equation}
where

\begin{equation}
G_{11}(t)= \dfrac{4 \eta  \left[2 \eta ^2+5 \eta +4 \beta  \eta  t+\beta  t (2 \beta  t+3)-3\right]}{\lambda +8 \pi}+\dfrac{\splitfrac{3 \left[\eta -2 (\eta +\beta  t)^2\right]^2+10 \eta  (\eta +\beta  t) (2 \eta +2 \beta  t-1)}{-2 \eta  (6 \eta +4 \beta  t-3)}}{\lambda +4 \pi},
\end{equation}

\begin{multline}
G_{21}(t)=\frac{2 \eta  \left[\eta(\eta -3)(1-2 \eta )^2+4 \beta ^4 t^4+2 \beta ^3 (8 \eta +3) t^3+2 \beta ^2 [2 \eta  (6 \eta -1)-3] t^2+\beta  \eta  (2 \eta -1) (8 \eta -9) t\right]}{\lambda +8 \pi }\\
+\dfrac{\splitfrac{6 \eta ^2 (2 \eta +2 \beta  t-1)^2-3 \left[2 (\eta +\beta  t)^2-\eta \right]^3+\eta  t (2 \eta +2 \beta  t-1) \left[2 (\eta +\beta  t)^2-\eta \right]}{+5 \eta  (\eta +\beta  t) (2 \eta +2 \beta  t-1) \left[2 (\eta +\beta  t)^2-\eta \right]}}{\lambda +4 \pi },
\end{multline}

\begin{multline}
F_{11}(t)=\eta  \left[\eta ^4 (\lambda +25.1327)+8.37758 \eta ^3+\eta^2  (-3.25 \lambda -60.7375)+\eta(1.5 \lambda +25.1327)\right]\\
 +\beta ^4 (1. \lambda +25.1327) t^4+\beta ^3 \eta  (4 \lambda +100.531) t^3+\beta ^2 \eta  t^2 [\eta  (6 \lambda +150.796)+8.37758]+\\
 \beta  \eta  t \left[\eta ^2 (4 \lambda +100.531)+16.7552 \eta -2.5 \lambda -50.2655\right],
\end{multline}

\begin{multline}
F_{21}(t)=\frac{8\eta \beta ^4 t^4+4 \eta \beta ^3 (8 \eta +3) t^3+4\eta \beta ^2 \left(12 \eta ^2-2 \eta -3\right) t^2+2 \eta^2\beta \left(16 \eta ^2-26 \eta +9\right)t+2 \eta(\eta^2 -3\eta) (1-2 \eta )^2}{(\lambda +8 \pi ) \left[\eta  (2 \eta -1)+2 \beta ^2 t^2+4 \beta  \eta  t\right]^2}-\\
\dfrac{\splitfrac{6 \eta ^2 (2 \eta +2 \beta  t-1)^2-3 \left[2 (\eta +\beta  t)^2-\eta \right]^3+\eta  t (2 \eta +2 \beta  t-1) \left[2 (\eta +\beta  t)^2-\eta \right]}{+5 \eta  (\eta +\beta  t) (2 \eta +2 \beta  t-1) \left[2 (\eta +\beta  t)^2-\eta \right]}}{(\lambda +4 \pi ) \left[\eta -2 (\eta +\beta  t)^2\right]^2}.
\end{multline}

\begin{figure}[h!]
\minipage{0.48\textwidth}
\includegraphics[width=75mm]{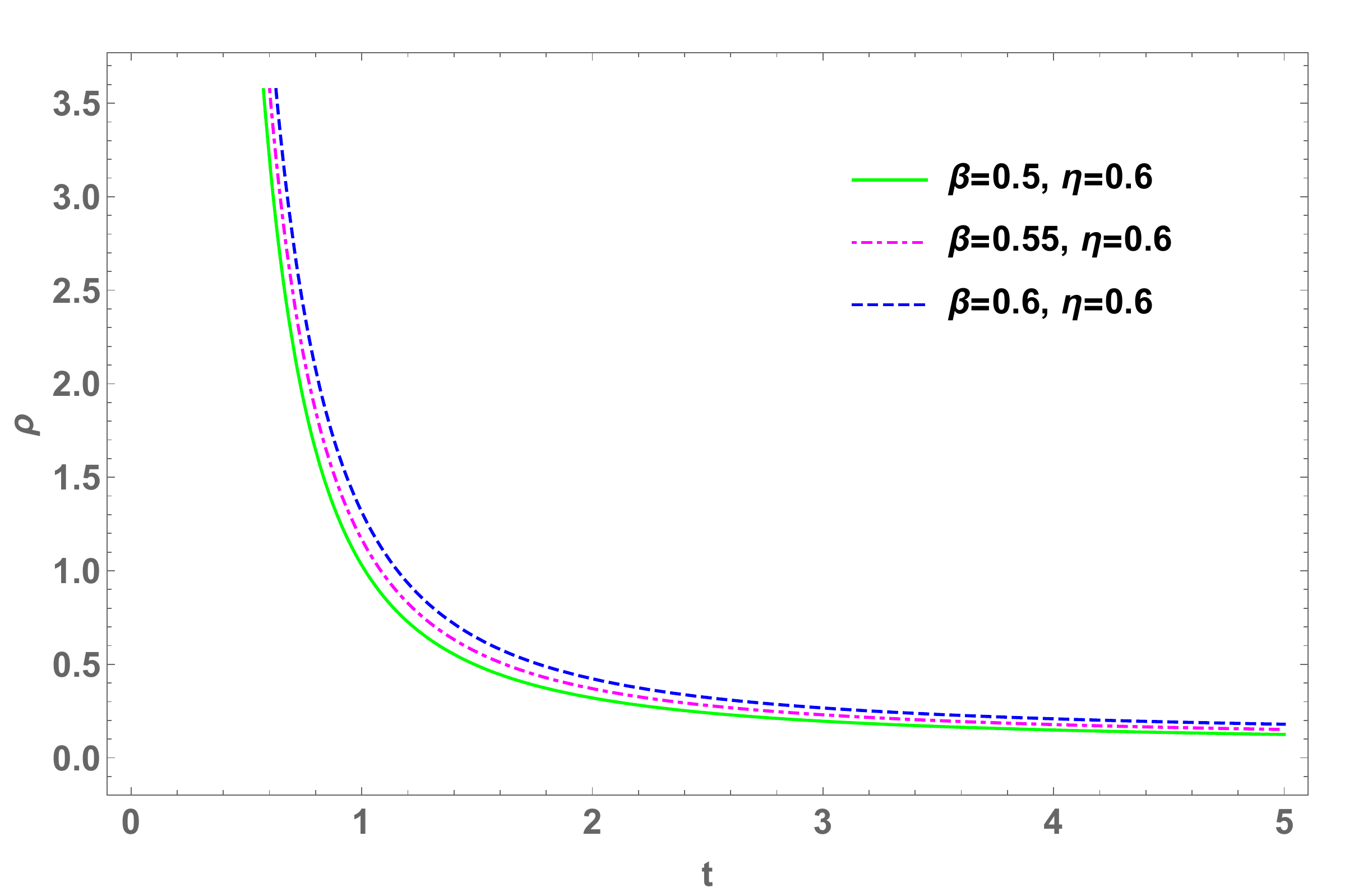}
  \caption{Variation of energy density against time with $\alpha=0.2$, $\mu=-1$, $\lambda=-8$.}\label{fig3}
\endminipage\hfill
\minipage{0.50\textwidth}
\includegraphics[width=75mm]{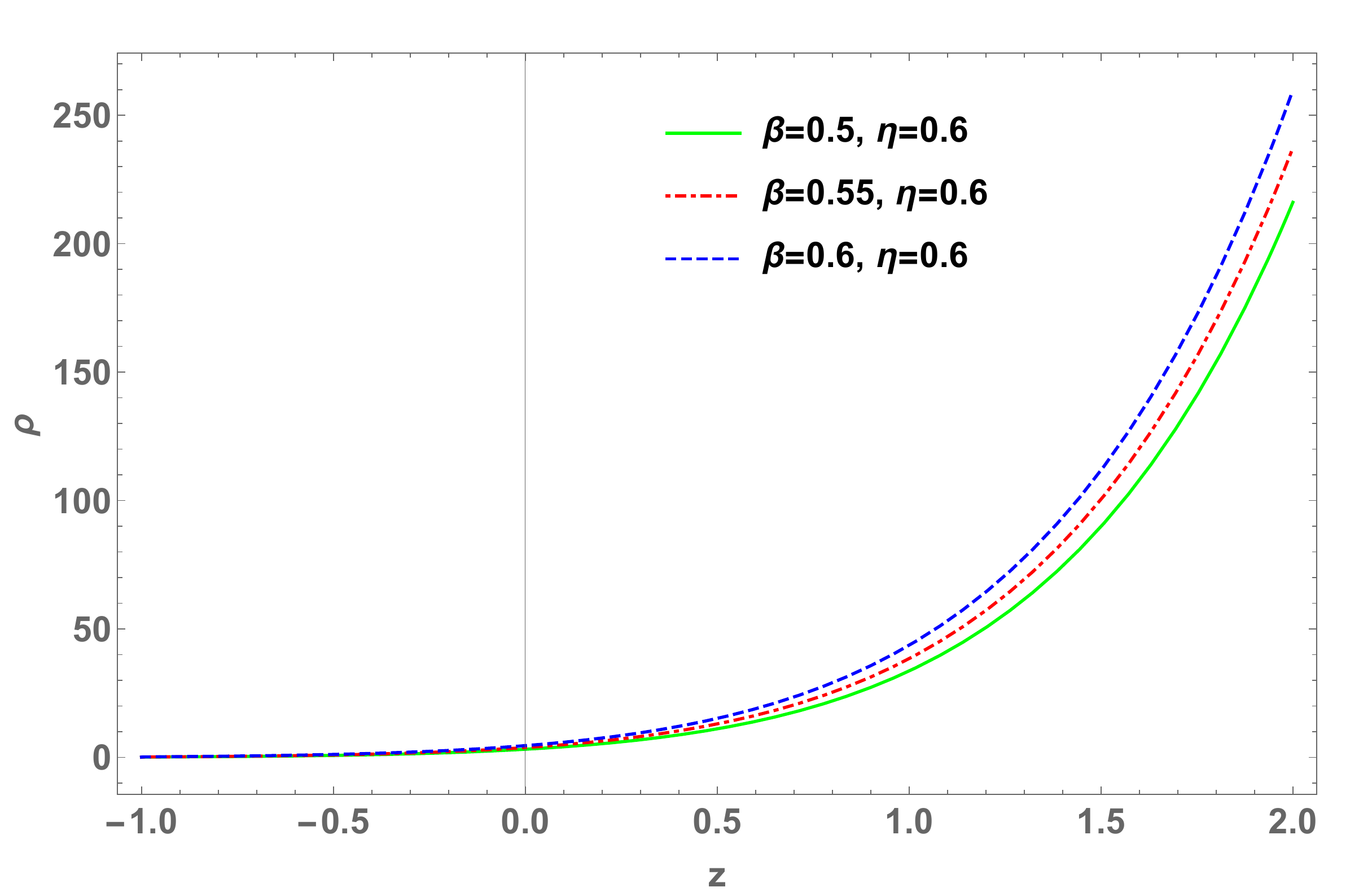}
  \caption{Variation of energy density against $z$ with $\alpha=0.2$, $\mu=-1$, $\lambda=-8$.}\label{fig3z}
\endminipage
\end{figure}
\begin{figure}[h!]
\minipage{0.48\textwidth}
\includegraphics[width=75mm]{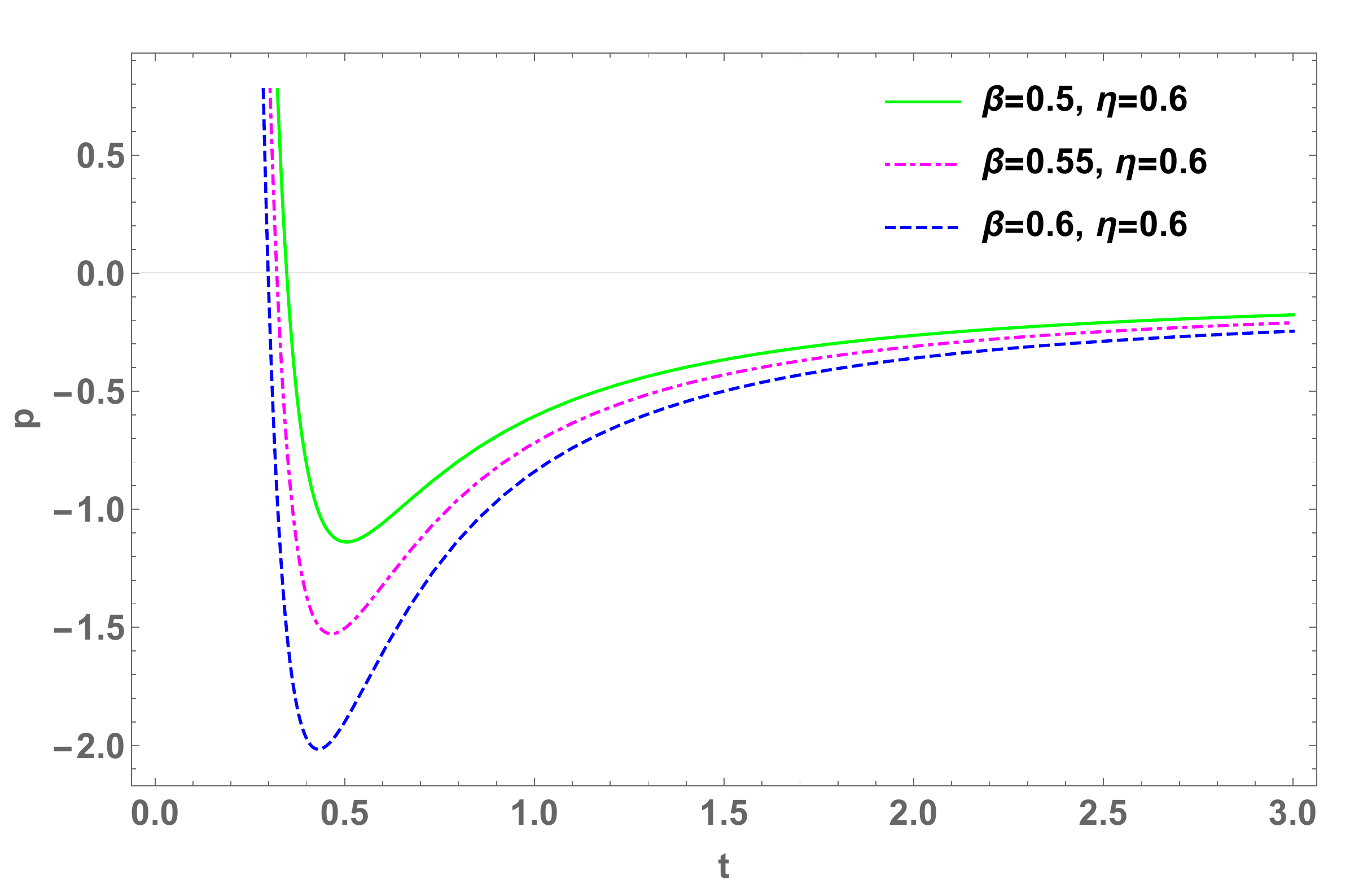}
  \caption{Variation of pressure against time with $\alpha=0.2$, $\mu=-1$, $\lambda=-8$.}\label{fig4}
\endminipage\hfill
\minipage{0.50\textwidth}
\includegraphics[width=75mm]{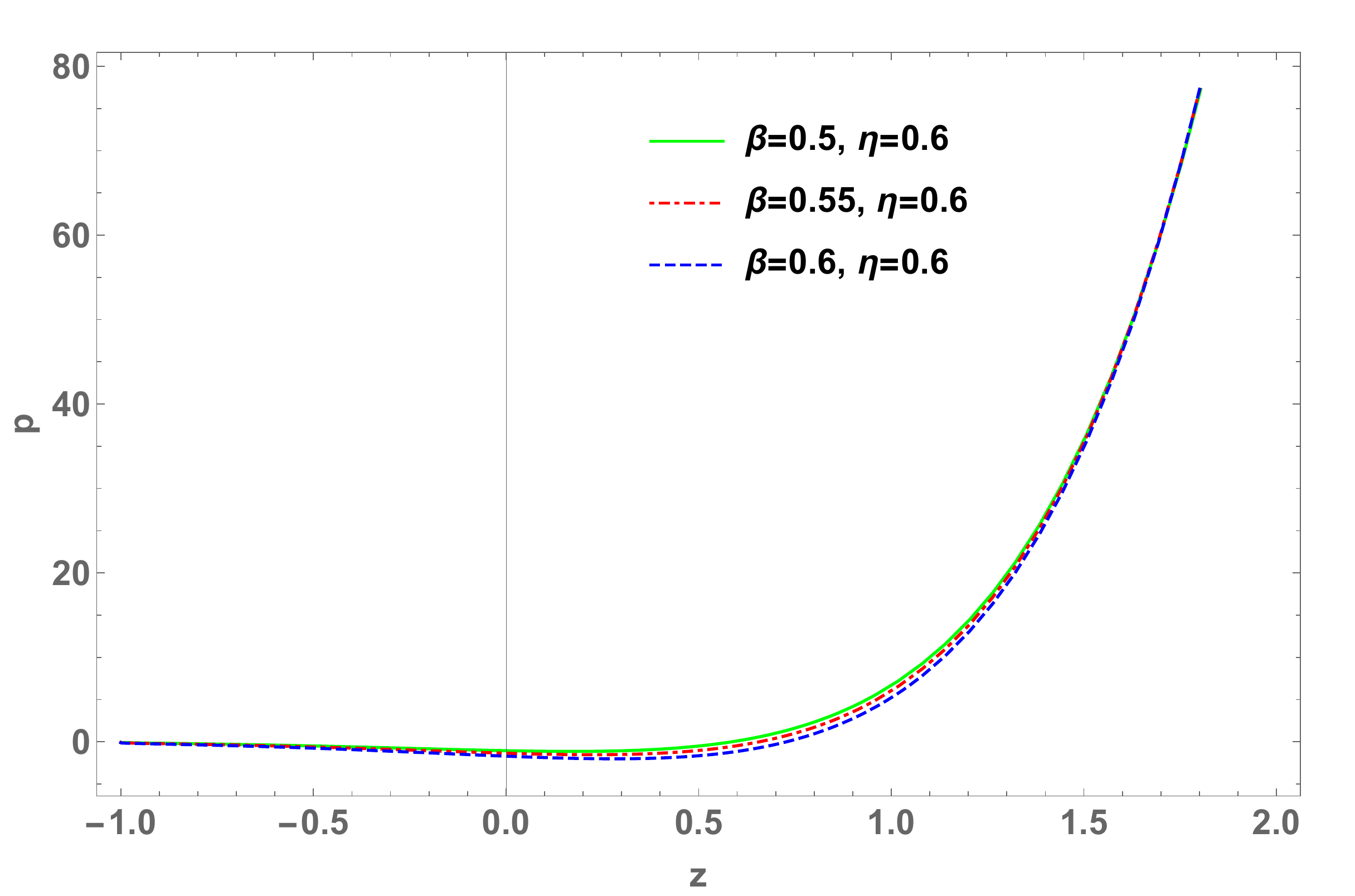}
  \caption{Variation of pressure against $z$ with $\alpha=0.2$, $\mu=-1$, $\lambda=-8$.}\label{fig4z}
\endminipage
\end{figure}
\begin{figure}[h!]
\minipage{0.48\textwidth}
\includegraphics[width=75mm]{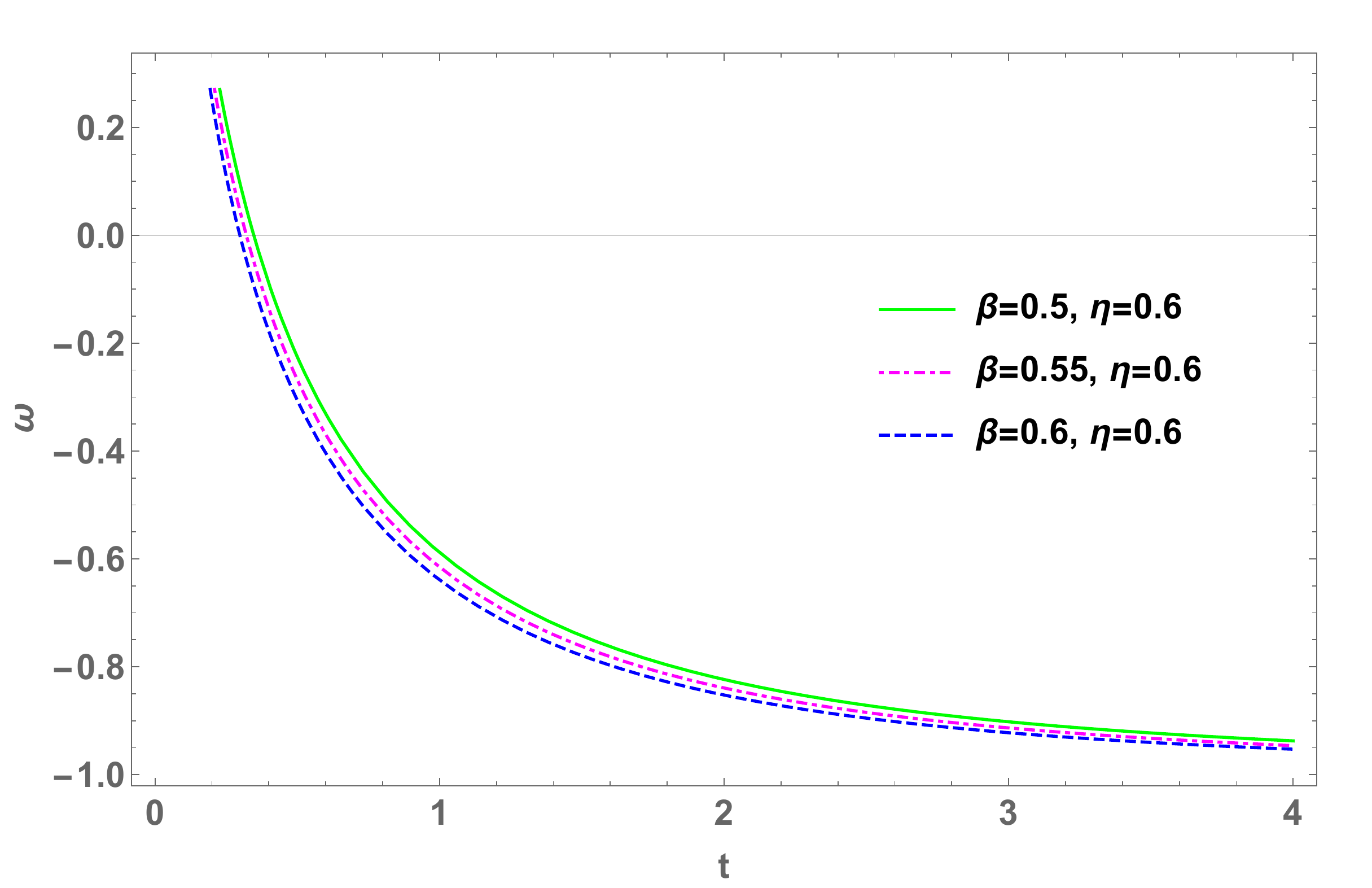}
  \caption{Variation of EoS Parameter against time with $\alpha=0.2$, $\mu=-1$, $\lambda=-8$.}\label{fig5}
\endminipage\hfill
\minipage{0.50\textwidth}
\includegraphics[width=75mm]{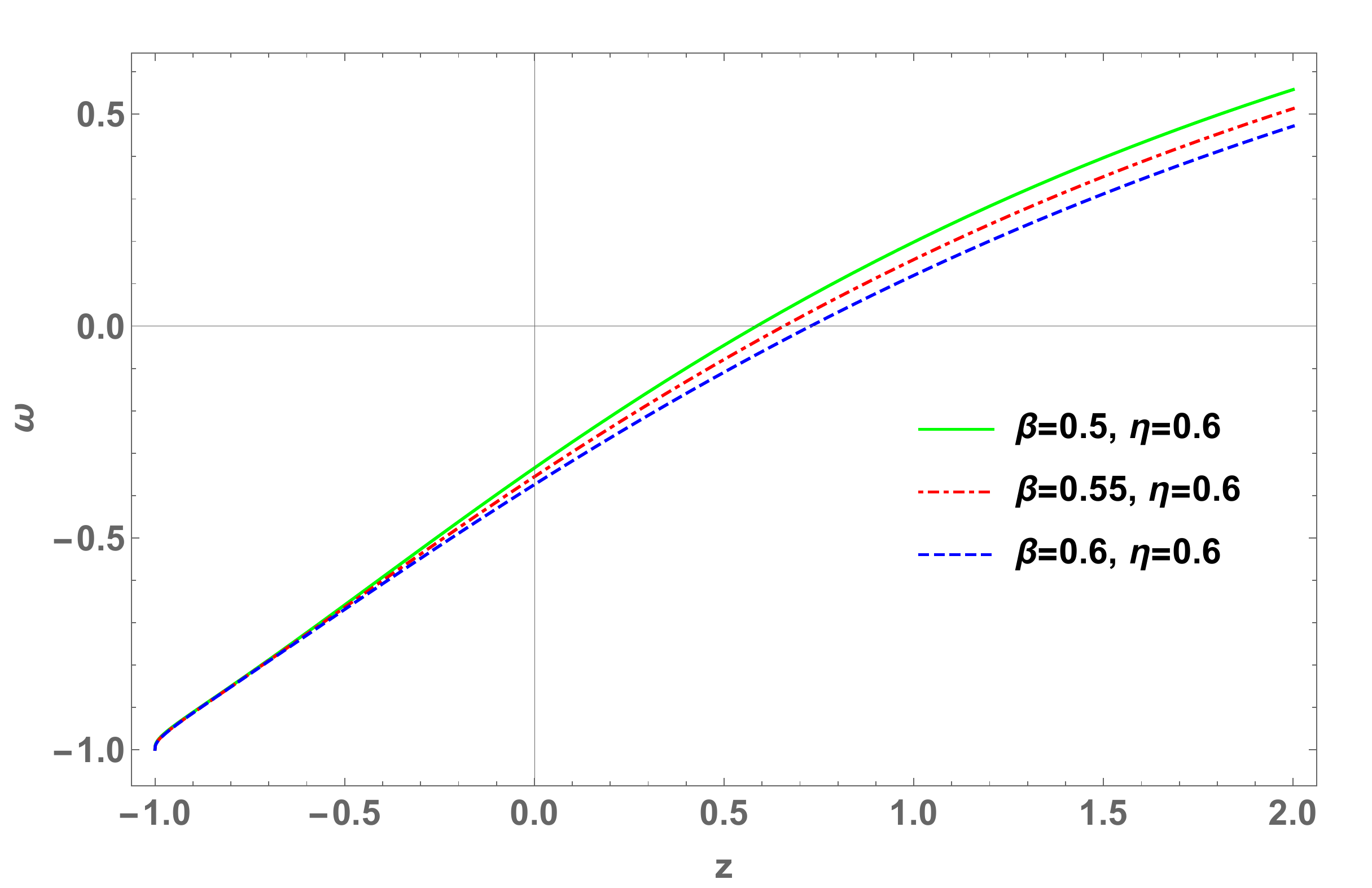}
  \caption{Variation of EoS Parameter against $z$ with $\alpha=0.2$, $\mu=-1$, $\lambda=-8$.}\label{fig5z}
\endminipage
\end{figure}

\subsection{The $f(R,T)=R+k\ln(\gamma R)+\lambda T$ Model}

By using $f(R,T)=R+k\ln(\gamma R)+\lambda T$ with Equation (\ref{e15}) in Equations (\ref{e12}-\ref{e14}), the analytical forms for $p$, $\rho$ and $\omega$ are written as

\begin{multline}\label{e20}
\rho=\frac{-0.0833333}{2 (\eta +\beta  t)^2-\eta } \left[k-\frac{ 12 (\eta +\beta  t)^2-6\eta}{t^2}\right] \left[\frac{2 \eta }{\lambda +8 \pi }+\frac{3 (\eta +\beta  t)^2-\eta }{\lambda +4 \pi }\right]+\\
\frac{\eta  k G_{12}(t)}{12 (\lambda +4 \pi ) (\lambda +8 \pi ) \left[2 (\eta +\beta  t)^2-\eta \right]^3} -\frac{k}{4\lambda +16 \pi} \left\{\log \left(-\frac{6 \gamma  \left[2 (\eta +\beta  t)^2-\eta \right]}{t^2}\right)-1\right\},
\end{multline}

\begin{multline}\label{e21}
p=\frac{-0.0833333 t^2}{2 (\eta +\beta  t)^2-\eta } \left[k-\frac{6 \left[2 (\eta +\beta  t)^2-\eta \right]}{t^2}\right] \left[\frac{2 \eta }{(\lambda +8 \pi ) t^2}-\frac{6 (\eta +\beta  t)^2-2 \eta}{(2 \lambda +8 \pi)t^2}\right]\\+
\frac{\eta  k F_{12}(t)}{12 (\lambda +4 \pi) (\lambda +8 \pi) \left[2 (\eta +\beta  t)^2-\eta \right]^3}
+\frac{k}{4\lambda +16 \pi} \left\{\log \left(-\frac{6 \gamma  \left[2 (\eta +\beta  t)^2-\eta \right]}{t^2}\right)-1\right\},
\end{multline}

\begin{equation}\label{e22}
\omega=\dfrac{\splitfrac{\frac{-0.0833333 t^2}{2 (\eta +\beta  t)^2-\eta } \left[k-\frac{6 \left[2 (\eta +\beta  t)^2-\eta \right]}{t^2}\right] \left[\frac{2 \eta }{(\lambda +8 \pi ) t^2}-\frac{6 (\eta +\beta  t)^2-2 \eta}{(2 \lambda +8 \pi)t^2}\right]\\+
\frac{\eta  k F_{12}(t)}{12 (\lambda +4 \pi) (\lambda +8 \pi) \left[2 (\eta +\beta  t)^2-\eta \right]^3}
}{+\frac{k}{4\lambda +16 \pi} \left\{\log \left(-\frac{6 \gamma  \left[2 (\eta +\beta  t)^2-\eta \right]}{t^2}\right)-1\right\}}}
{\splitfrac{\frac{-0.0833333}{2 (\eta +\beta  t)^2-\eta } \left[k-\frac{ 12 (\eta +\beta  t)^2-6\eta}{t^2}\right] \left[\frac{2 \eta }{\lambda +8 \pi }+\frac{3 (\eta +\beta  t)^2-\eta }{\lambda +4 \pi }\right]+\\
\frac{\eta  k G_{12}(t)}{12 (\lambda +4 \pi ) (\lambda +8 \pi ) \left[2 (\eta +\beta  t)^2-\eta \right]^3} }{-\frac{k}{4\lambda +16 \pi} \left\{\log \left(-\frac{6 \gamma  \left[2 (\eta +\beta  t)^2-\eta \right]}{t^2}\right)-1\right\}}},
\end{equation}
where

\begin{multline}
G_{12}(t)=\lambda  \biggl[-\eta  (7 \eta -1)(1-2 \eta )^2 -28 \beta ^4 t^4+2 \beta ^3 (3-56 \eta ) t^3+2 \beta ^2 \left(-84 \eta ^2+22 \eta +3\right) t^2
-7 \beta  \eta  (2 \eta -1) (8 \eta -1) t\biggr]\\-48 \pi  (\eta +\beta  t) (2 \eta +2 \beta  t-1) \left[2 (\eta +\beta  t)^2-\eta \right],
\end{multline}
\begin{multline}
F_{12}(t)=4 \beta ^4 t^4 (3 \lambda +32 \pi)+2 \beta ^3 t^3 [3 \lambda (8 \eta -5) +32\pi (8 \eta -3)]+2 \beta ^2 t^2 \{6 \eta\lambda (6 \eta -5) +9 \lambda\\
 +16\pi[2 \eta (12 \eta -7)+3]\}+\beta  \eta  (2 \eta -1) (8 \eta -1)(3 \lambda +32 \pi) t+(1-2 \eta )^2 [3 \eta \lambda (\eta +1) +16 \eta \pi (2 \eta +1)].
\end{multline}

\begin{figure}[h!]
\minipage{0.48\textwidth}
\includegraphics[width=75mm]{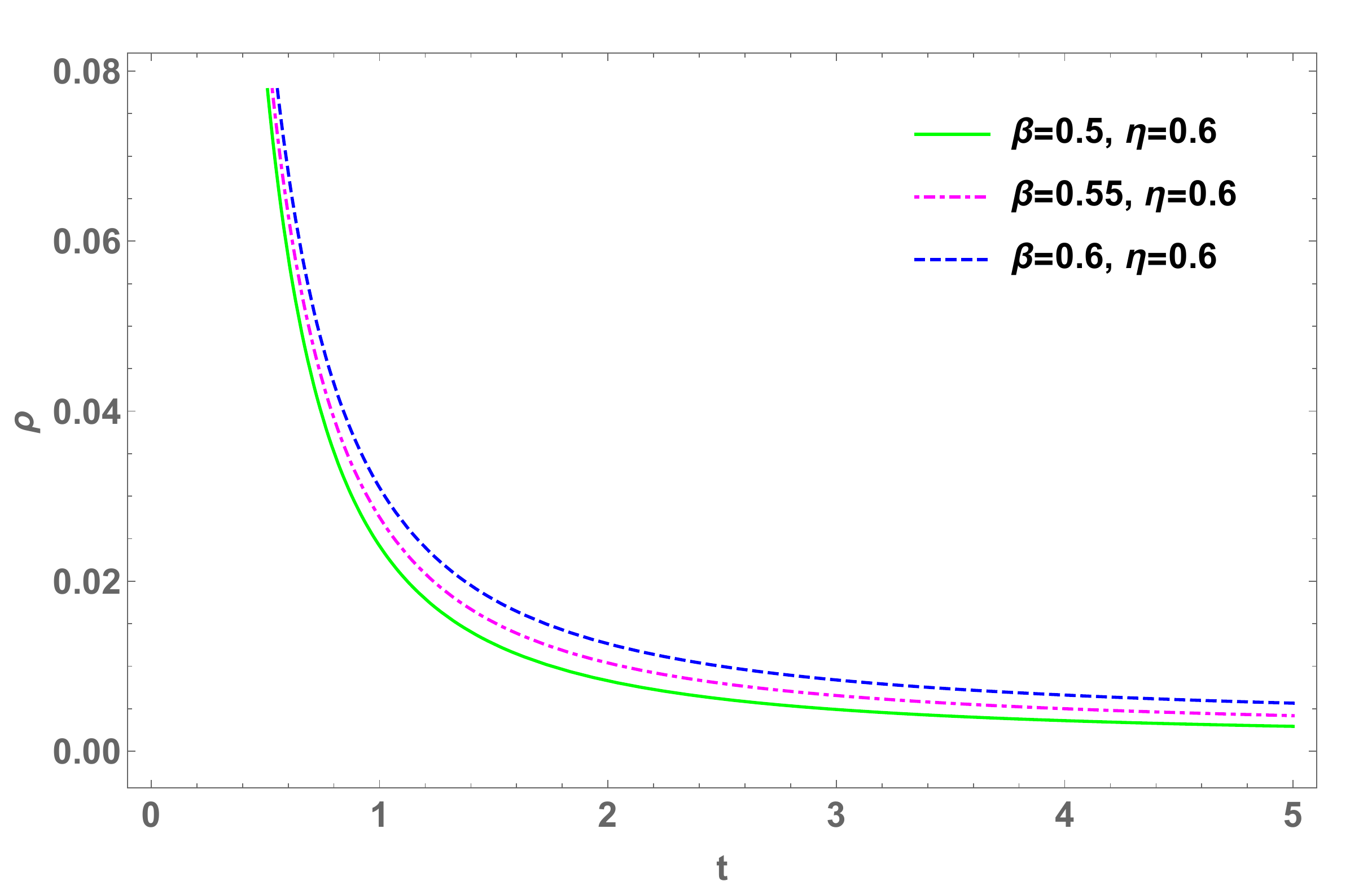}
  \caption{Variation of energy density against time with $k=1$, $\gamma=-2$, $\lambda=35$.}\label{fig6}
\endminipage\hfill
\minipage{0.50\textwidth}
  \includegraphics[width=75mm]{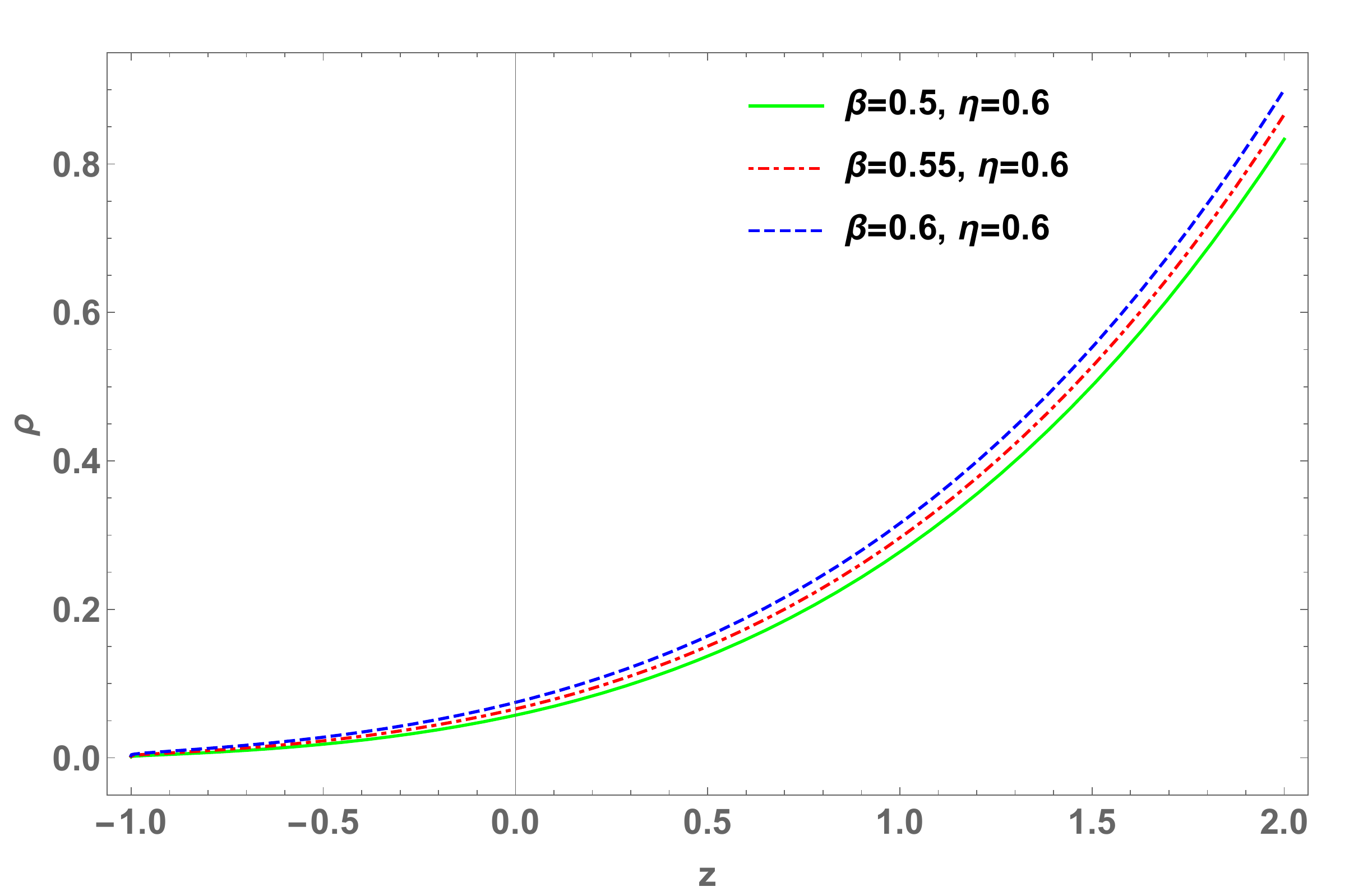}
  \caption{Variation of energy density against $z$ with $k=1$, $\gamma=-2$, $\lambda=35$.}\label{fig6z}
\endminipage
\end{figure}

\begin{figure}[h!]
\minipage{0.48\textwidth}
\includegraphics[width=75mm]{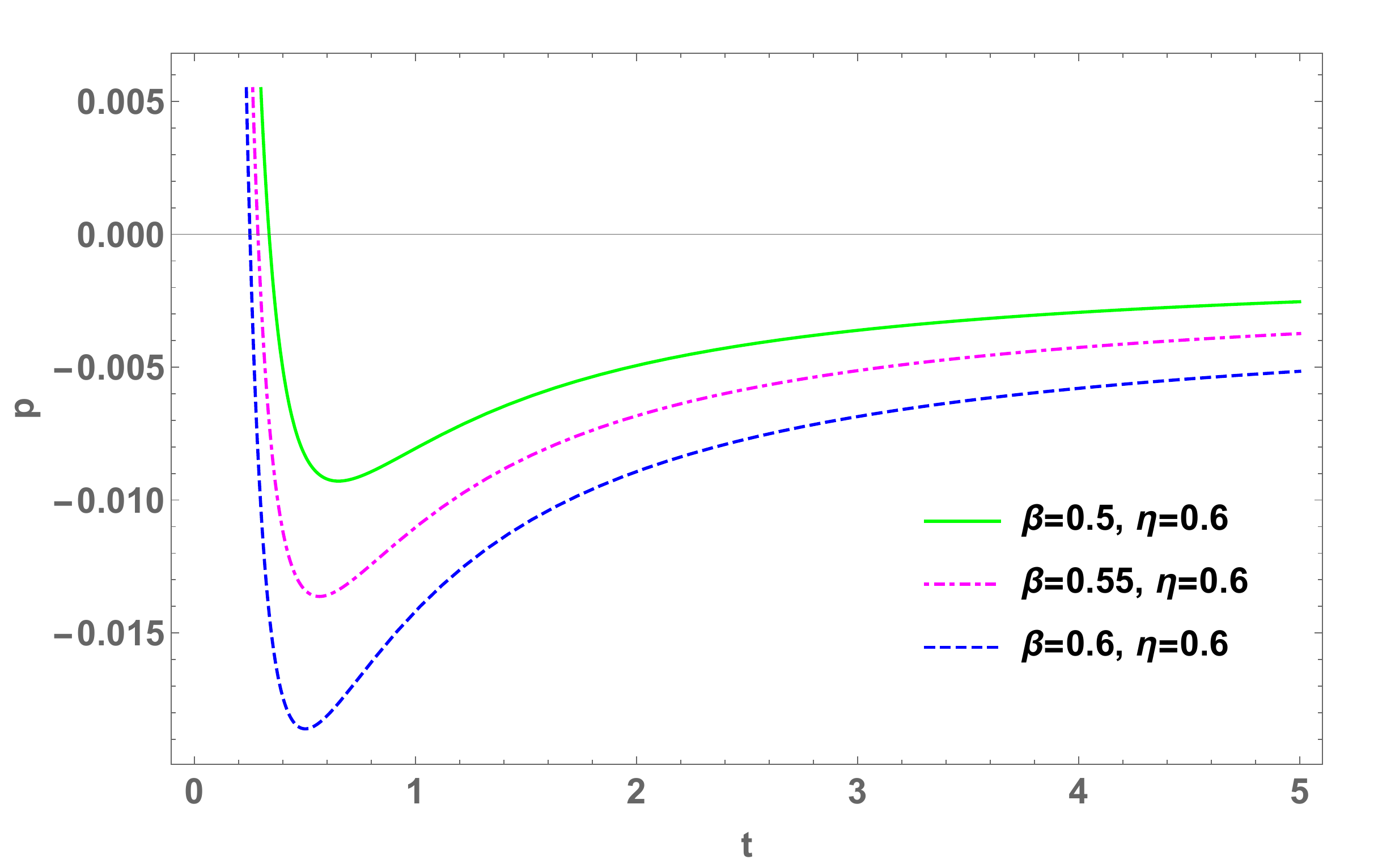}
  \caption{Variation of pressure against time with $k=1$, $\gamma=-2$, $\lambda=35$.}\label{fig7}
\endminipage\hfill
\minipage{0.50\textwidth}
  \includegraphics[width=75mm]{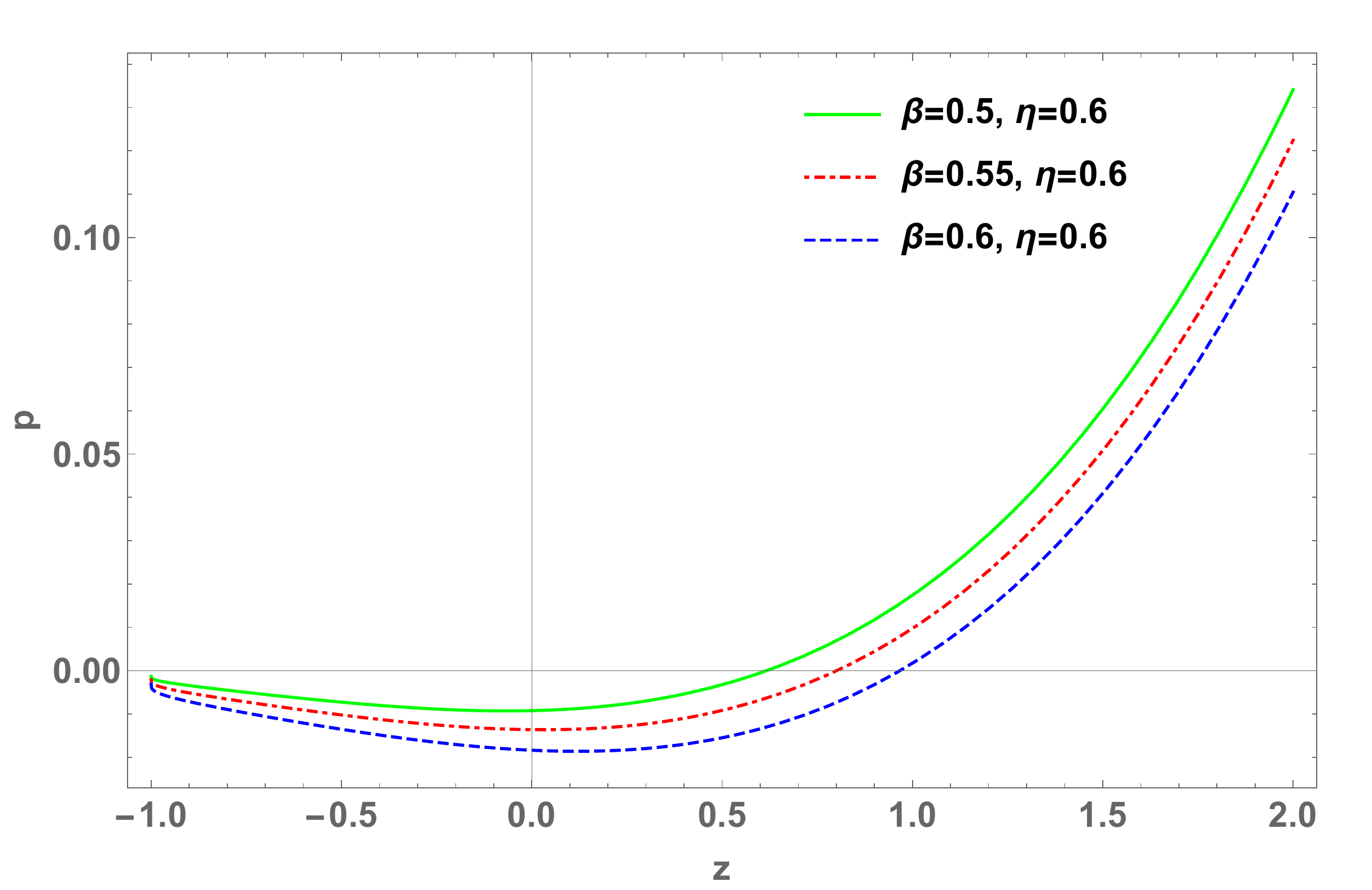}
  \caption{Variation of pressure against $z$ with $k=1$, $\gamma=-2$, $\lambda=35$.}\label{fig7z}
\endminipage
\end{figure}

\begin{figure}[h!]
\minipage{0.48\textwidth}
\includegraphics[width=75mm]{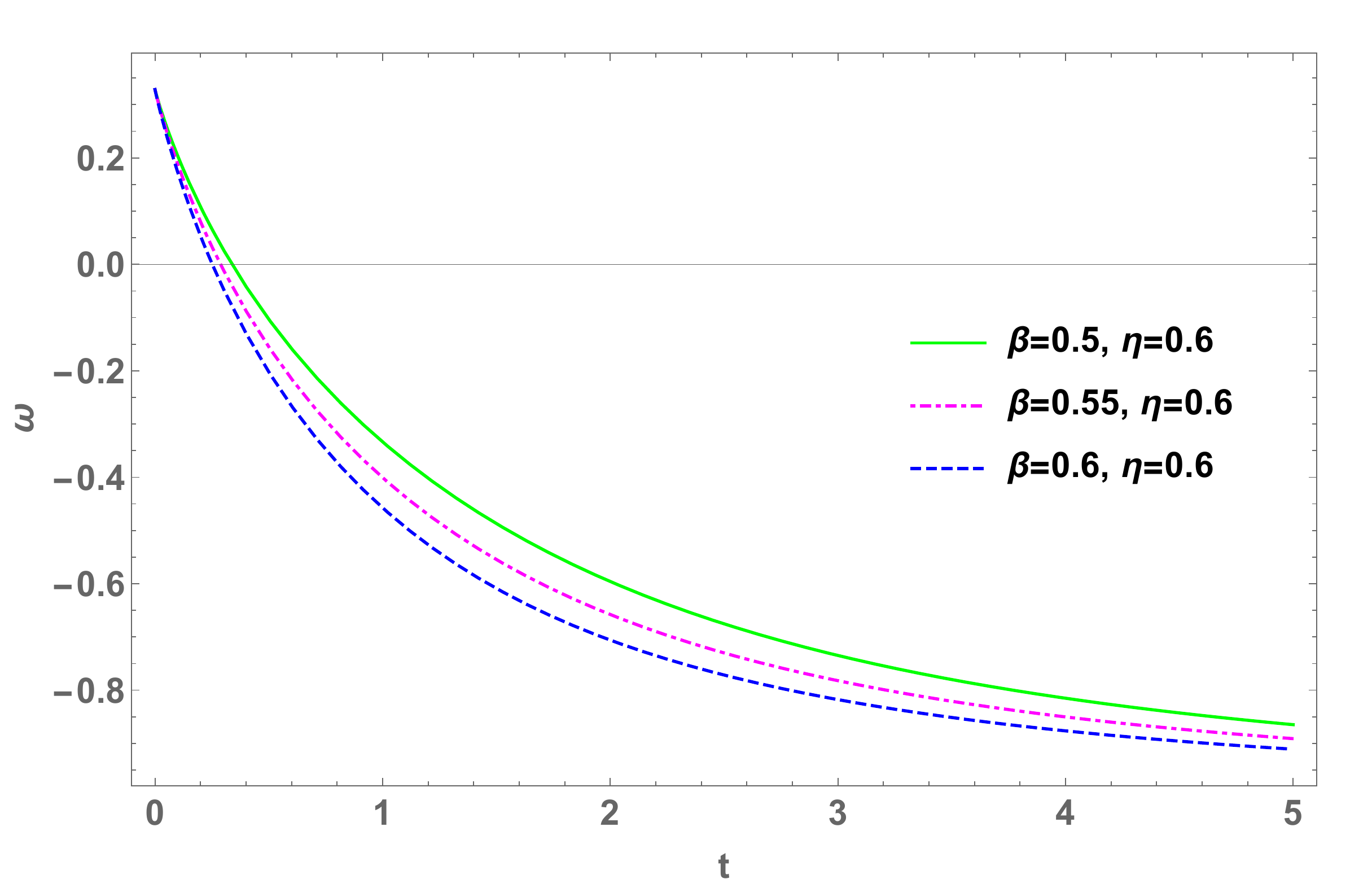}
  \caption{Variation of EoS Parameter against time with $k=1$, $\gamma=-2$, $\lambda=35$.}\label{fig8}
\endminipage\hfill
\minipage{0.50\textwidth}
  \includegraphics[width=75mm]{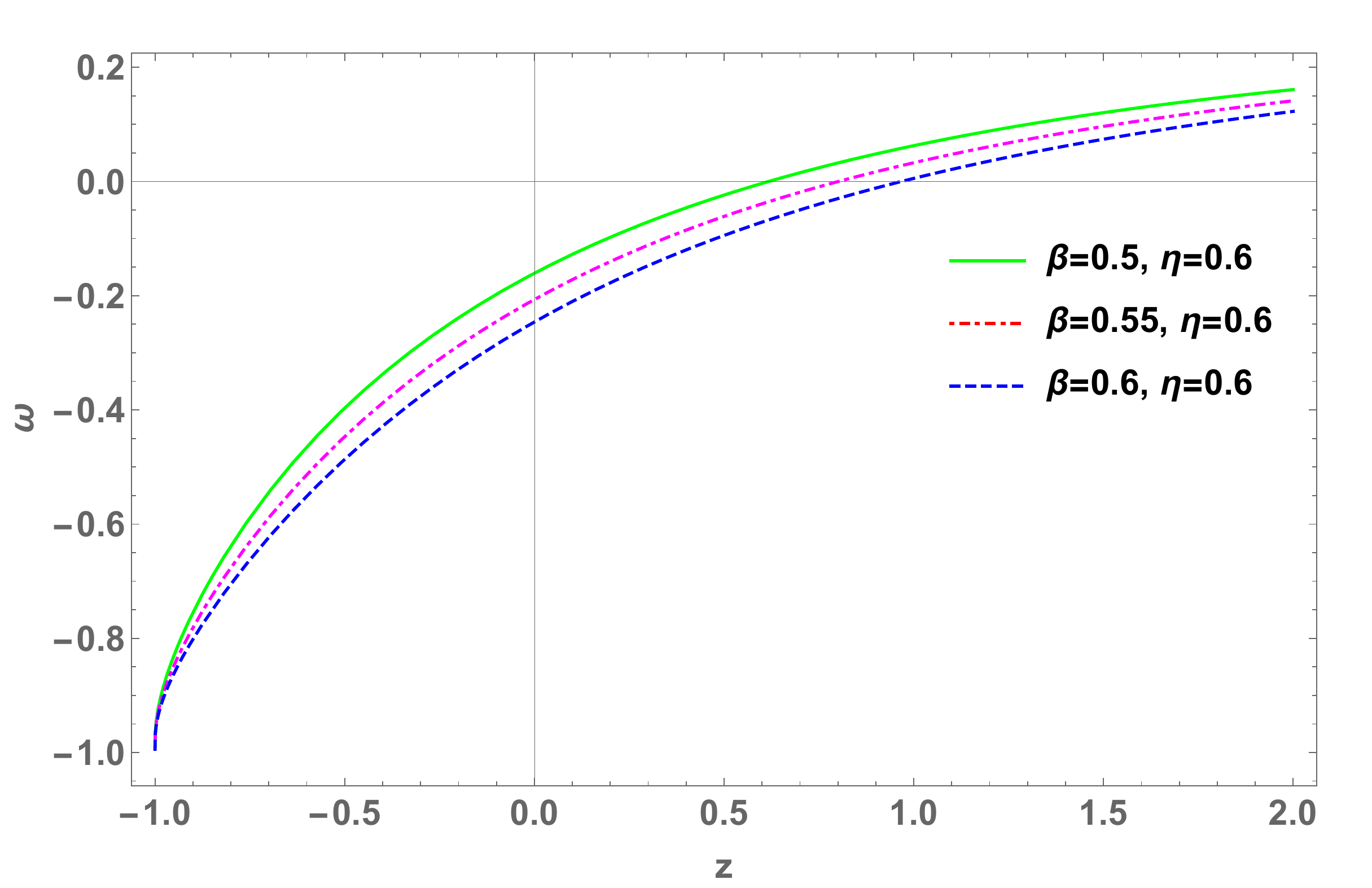}
  \caption{Variation of EoS Parameter against $z$ with $k=1$, $\gamma=-2$, $\lambda=35$.}\label{fig8z}
\endminipage
\end{figure}

\subsection{The $f(R,T)=R+m e^{-nR}+\lambda T$ Model}

By taking $f(R,T)=R+m e^{-nR}+\lambda T$ and Equation (\ref{e15}) in Equations (\ref{e12}-\ref{e14}), the analytical forms for $p$, $\rho$ and $\omega$ are expressed as

\begin{multline}\label{e23}
\rho=\left\{1-m n e^{\frac{6 n \left[2 (\eta +\beta  t)^2-\eta \right]}{t^2}}\right\}\left[\frac{ \eta }{t^2 (\lambda +8 \pi)}-\frac{3 (\eta +\beta  t)^2- \eta}{(2 \lambda +8 \pi)t^2}\right]
+\frac{3 \eta  m n^2 e^{\frac{6 n \left[2 (\eta +\beta  t)^2-\eta \right]}{t^2}}G_{13}(t)}{t^6} \\
-\frac{m e^{\frac{6 n \left[2 (\eta +\beta  t)^2-\eta \right]}{t^2}}}{4 (\lambda +4 \pi )} \left\{1-\frac{6 n \left[2 (\eta +\beta  t)^2-\eta \right]}{t^2}\right\},
\end{multline}

\begin{multline}\label{e24}
p=\left\{1-m n e^{\frac{6 n \left[2 (\eta +\beta  t)^2-\eta \right]}{t^2}}\right\}\left[\frac{ \eta }{(\lambda +8 \pi ) t^2}-\frac{3 (\eta +\beta  t)^2- \eta}{(2 \lambda +8 \pi)t^2 }\right] +\frac{3 \eta  m n^2 e^{\frac{6 n \left[2 (\eta +\beta  t)^2-\eta \right]}{t^2}}F_{13}(t) }{t^6}\\
+\frac{m e^{\frac{6 n \left[2 (\eta +\beta  t)^2-\eta \right]}{t^2}}}{4 (\lambda +4 \pi)} \left\{1-\frac{6 n \left[2 (\eta +\beta  t)^2-\eta \right]}{t^2}\right\},
\end{multline}

\begin{equation}\label{e25}
\omega=\dfrac{\left\{1-m n e^{\frac{6 n \left[2 (\eta +\beta  t)^2-\eta \right]}{t^2}}\right\}\left[\frac{ \eta }{(\lambda +8 \pi ) t^2}-\frac{3 (\eta +\beta  t)^2- \eta}{(2 \lambda +8 \pi)t^2 }\right] +\frac{3 \eta  m n^2 e^{\frac{6 n \left[2 (\eta +\beta  t)^2-\eta \right]}{t^2}}F_{13}(t) }{t^6}\\
+\frac{m e^{\frac{6 n \left[2 (\eta +\beta  t)^2-\eta \right]}{t^2}}}{4 (\lambda +4 \pi)} \left\{1-\frac{6 n \left[2 (\eta +\beta  t)^2-\eta \right]}{t^2}\right\}}
{\left\{1-m n e^{\frac{6 n \left[2 (\eta +\beta  t)^2-\eta \right]}{t^2}}\right\}\left[\frac{ \eta }{t^2 (\lambda +8 \pi)}-\frac{3 (\eta +\beta  t)^2- \eta}{(2 \lambda +8 \pi)t^2}\right]
+\frac{3 \eta  m n^2 e^{\frac{6 n \left[2 (\eta +\beta  t)^2-\eta \right]}{t^2}}G_{13}(t)}{t^6} \\
-\frac{m e^{\frac{6 n \left[2 (\eta +\beta  t)^2-\eta \right]}{t^2}}}{4 (\lambda +4 \pi )} \left\{1-\frac{6 n \left[2 (\eta +\beta  t)^2-\eta \right]}{t^2}\right\}},
\end{equation}
where

\begin{multline}
G_{13}(t)=\frac{12 \eta  n (2 \eta +2 \beta  t-1)^2-5 t^2 (\eta +\beta  t) (2 \eta +2 \beta  t-1)+t^2 (6 \eta +4 \beta  t-3)}{\lambda +4 \pi}\\
+\frac{2 \left[12 \eta  n (2 \eta +2 \beta  t-1)^2+t^2 (\eta +\beta  t) (2 \eta +2 \beta  t-1)+t^2 (6 \eta +4 \beta  t-3)\right]}{\lambda +8 \pi},
\end{multline}

\begin{multline}
F_{13}(t)=\frac{12 \eta  n (2 \eta +2 \beta  t-1)^2-5 t^2 (\eta +\beta  t) (2 \eta +2 \beta  t-1)+t^2 (6 \eta +4 \beta  t-3)}{\lambda +4 \pi}\\
+\frac{2 \left[12 \eta  n (2 \eta +2 \beta  t-1)^2+t^2 (\eta +\beta  t) (2 \eta +2 \beta  t-1)+t^2 (6 \eta +4 \beta  t-3)\right]}{\lambda +8 \pi}.
\end{multline}

\begin{figure}[h!]
\minipage{0.48\textwidth}
 \includegraphics[width=75mm]{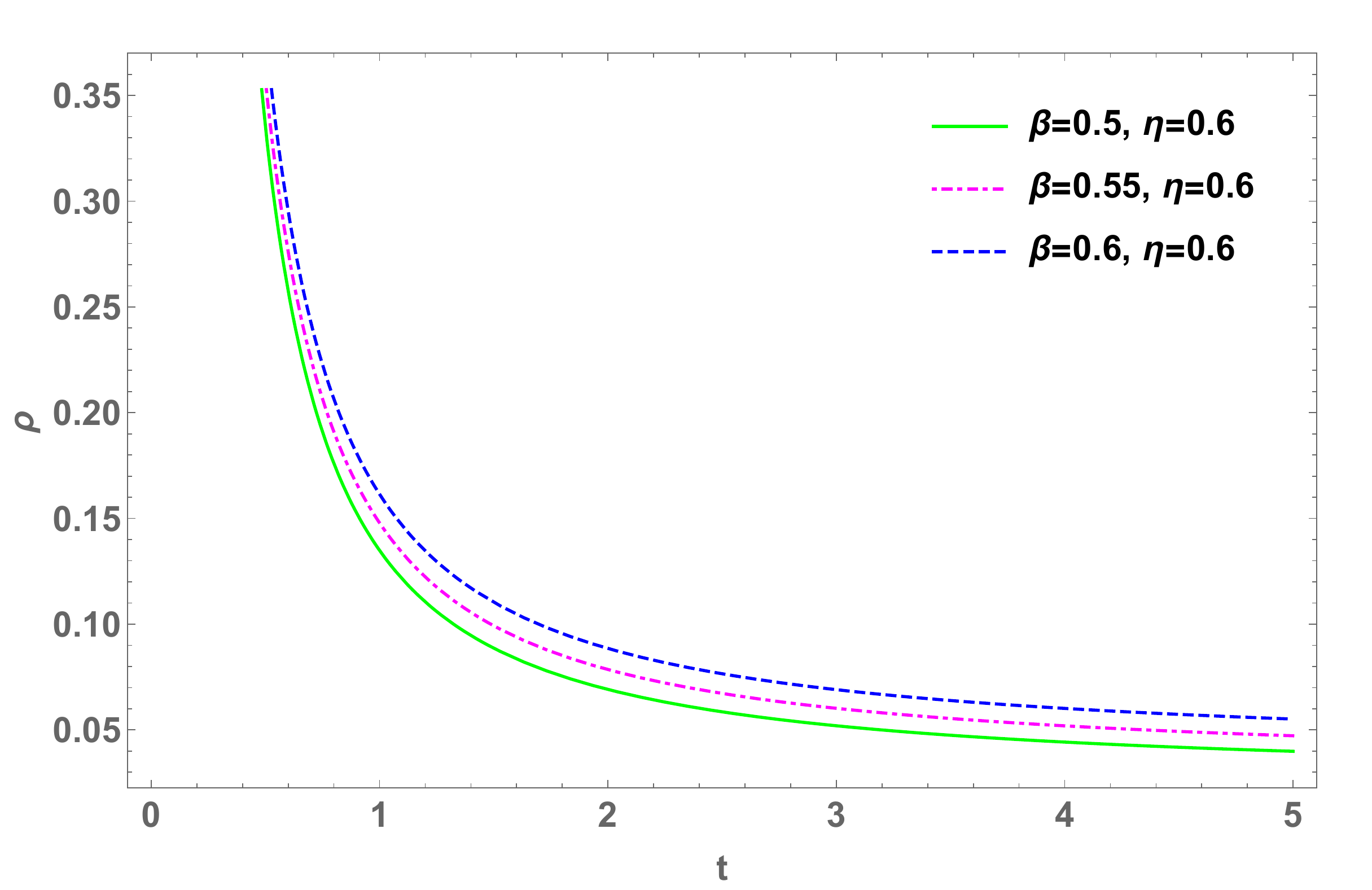}
  \caption{Variation of energy density against time with $m=0.2$, $n=0.05$, $\lambda=0.5$.}\label{fig9}
\endminipage\hfill
\minipage{0.50\textwidth}
 \includegraphics[width=75mm]{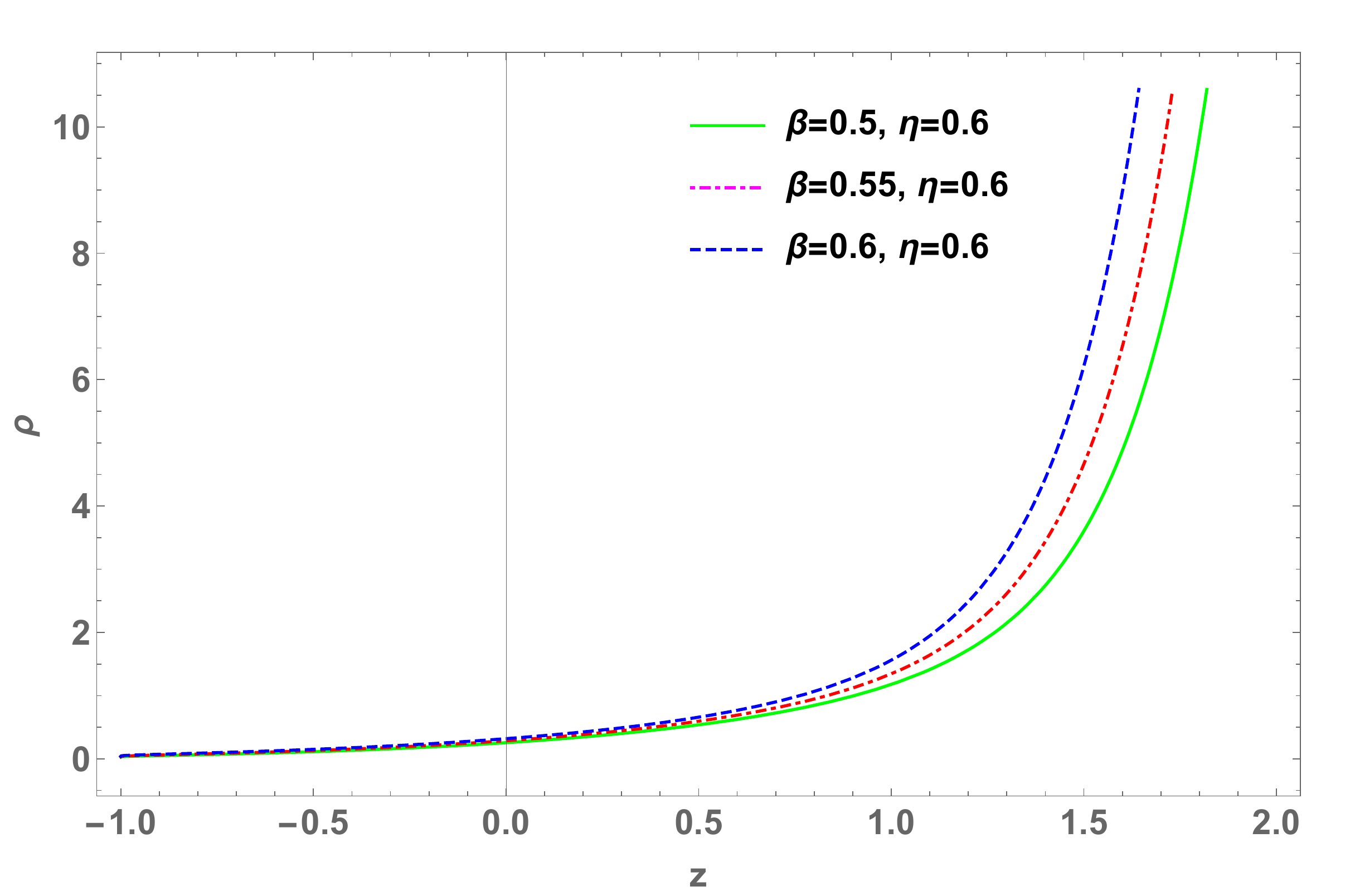}
  \caption{Variation of energy density against $z$ with $m=0.2$, $n=0.05$, $\lambda=0.5$.}\label{fig9z}
\endminipage
\end{figure}

\begin{figure}[h!]
\minipage{0.48\textwidth}
 \includegraphics[width=75mm]{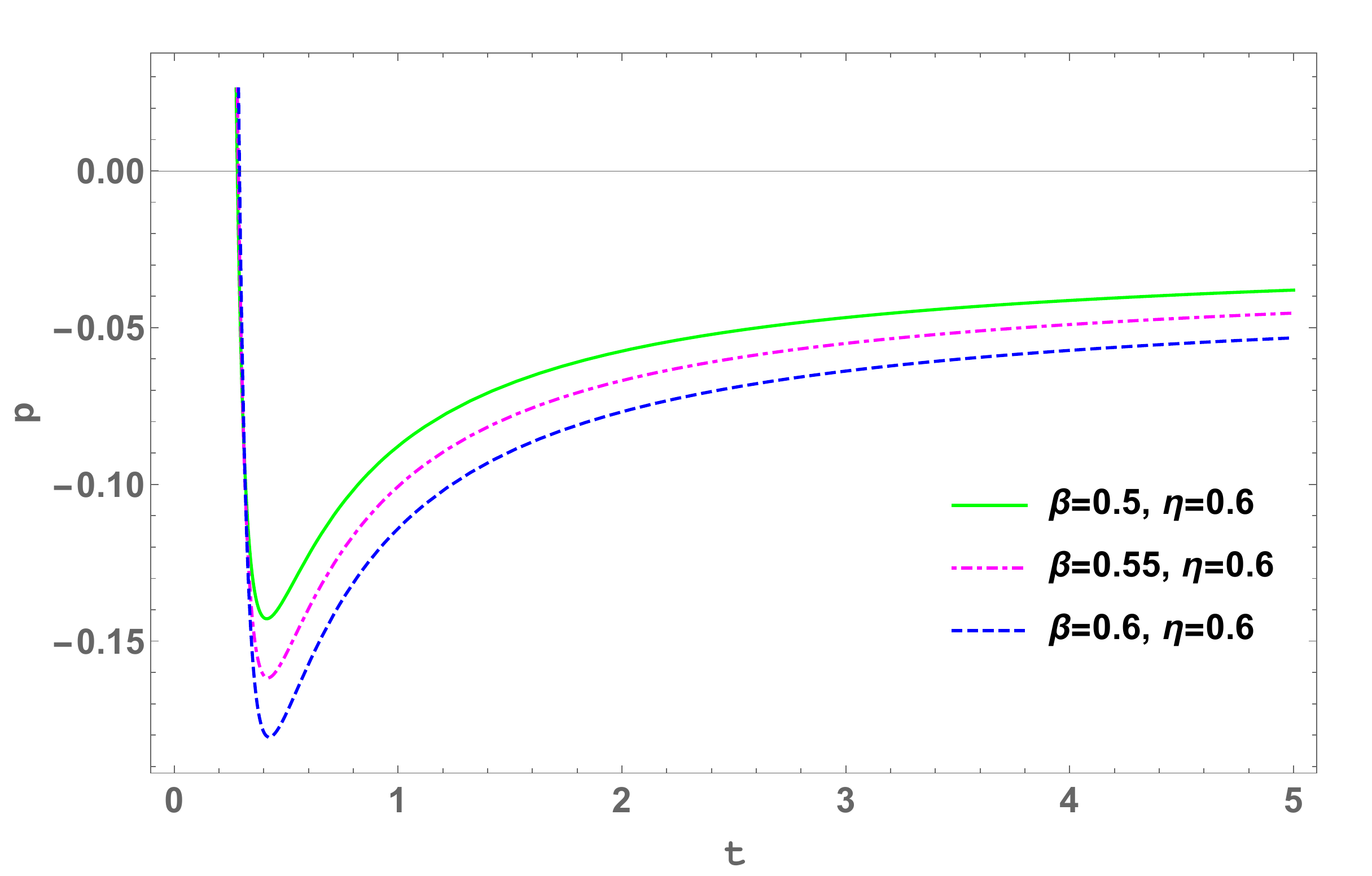}
  \caption{Variation of pressure against time with $m=0.2$, $n=0.05$, $\lambda=0.5$.}\label{fig10}
\endminipage\hfill
\minipage{0.50\textwidth}
 \includegraphics[width=75mm]{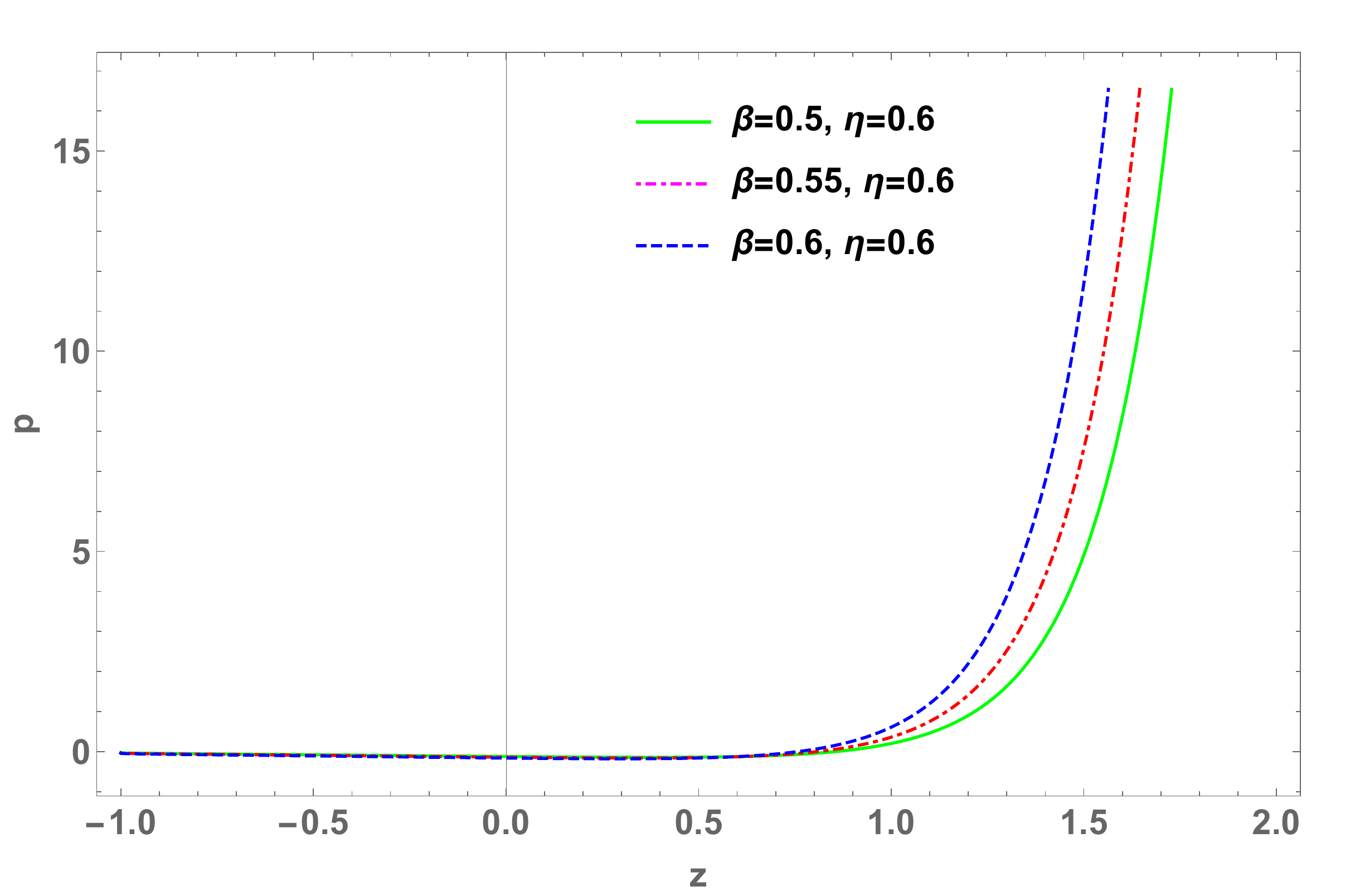}
  \caption{Variation of pressure against $z$ with $m=0.2$, $n=0.05$, $\lambda=0.5$.}\label{fig10z}
\endminipage
\end{figure}

\begin{figure}[h!]
\minipage{0.48\textwidth}
 \includegraphics[width=75mm]{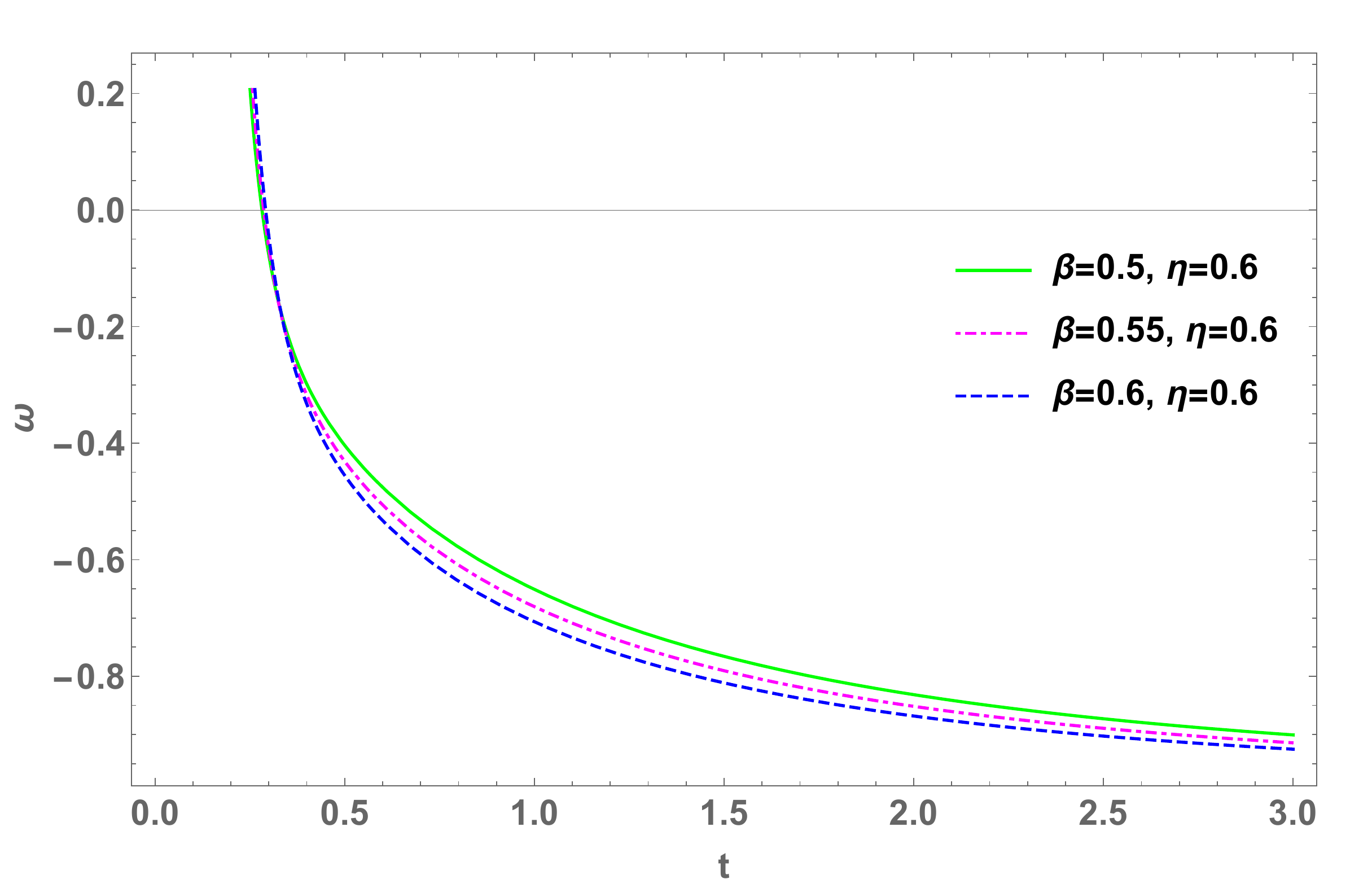}
  \caption{Variation of EoS Parameter against time with $m=0.2$, $n=0.05$, $\lambda=0.5$.}\label{fig11}
\endminipage\hfill
\minipage{0.50\textwidth}
 \includegraphics[width=75mm]{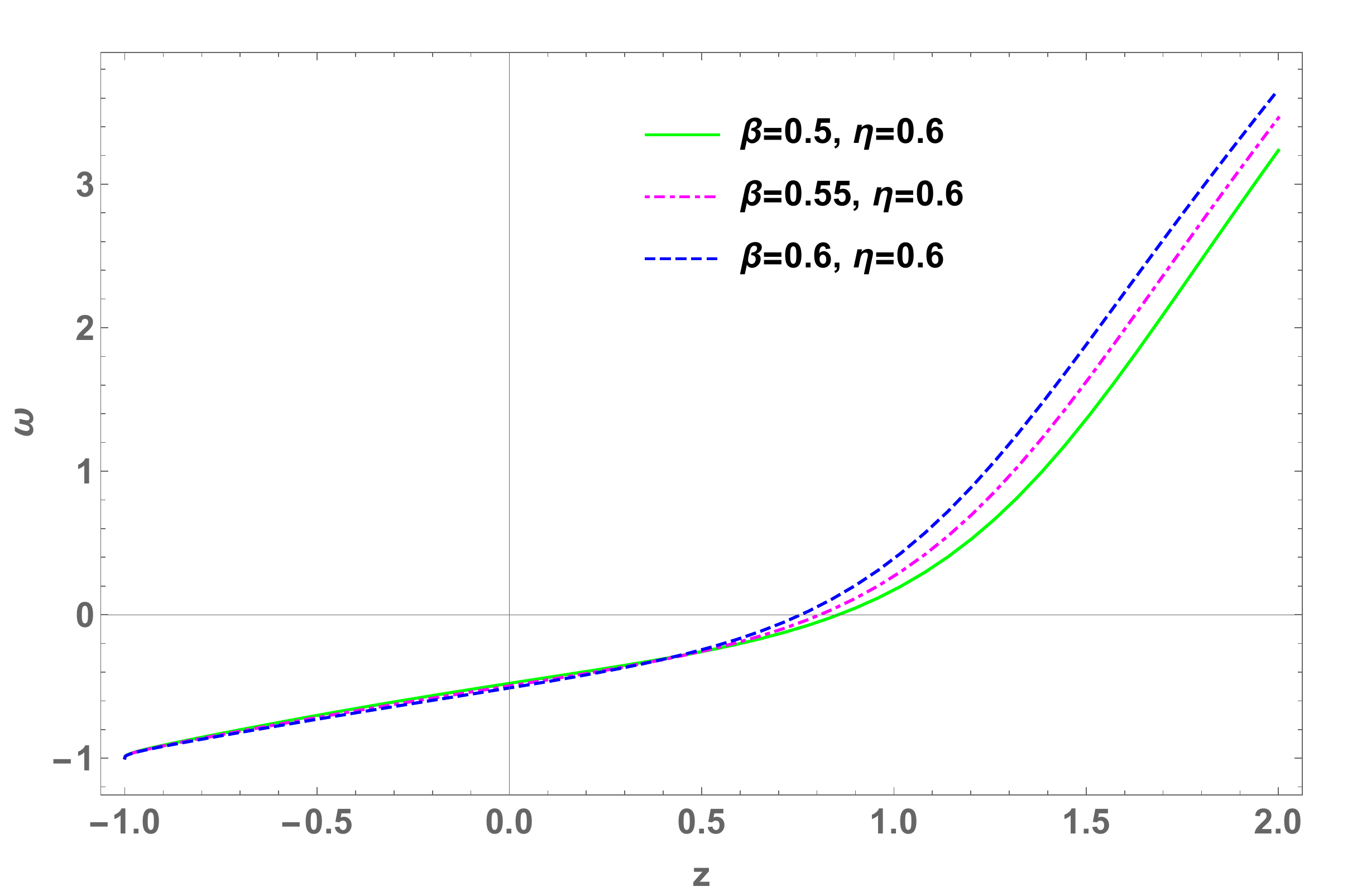}
  \caption{Variation of EoS Parameter against $z$ with $m=0.2$, $n=0.05$, $\lambda=0.5$.}\label{fig11z}
\endminipage
\end{figure}

\section{Om Diagnostic Analysis}

In the literature, state finder parameters and Om diagnostic analysis are used to differentiate dark energy models \cite{Sahni/2008}. In order to understand the cosmological models, the Hubble, deceleration and EoS parameters play an important role. It is known from the literature that dark energy models produce a positive Hubble parameter and a negative deceleration parameter. So $H$ and $q$ cannot be used to differentiate effectively between different dark energy models. Thus Om diagnostic analysis plays a crucial role for such analysis. The Om diagnosis has also been applied to Galileons models \cite{Jamil/2013,Fromont/2013}. The Om$(z)$ parameter for spatially flat universe is given by \cite{Sahni/2008,Zunckel/2008}

\begin{equation}
Om(z)=\frac{\left[\frac{H(z)}{H_0}\right]^2-1}{(1+z)^3-1}.
\end{equation}
Here, $H_0$ is the present value of the Hubble parameter. One can observe that the Om$(z)$ parameter involves first derivatives of the scale factor, so Om diagnosis is a simpler diagnostic than the state finder diagnosis. The positive, negative and zero values of Om$(z)$ represent the phantom ($\omega<-1$),  quintessence ($\omega>-1$) and $\Lambda$CDM dark energy models, respectively \cite{Shahalam/2015}.

In our discussed models, the Om$(z)$ parameter takes the form

\begin{equation}
Om(z)=\frac{(\beta^2-H_0^2)W^2\left[\frac{\beta  \left(\frac{1}{z+1}\right)^{1/\eta }}{\eta }\right]+2\beta^2W\left[\frac{\beta  \left(\frac{1}{z+1}\right)^{1/\eta }}{\eta }\right]+\beta^2}{W^2\left[\frac{\beta  \left(\frac{1}{z+1}\right)^{1/\eta }}{\eta }\right]H_0^2z(3+3z+z^2)},
\end{equation}
and its behaviour can be seen in the Fig. 20.

\begin{figure}[h!]
\centering
\includegraphics[width=75mm]{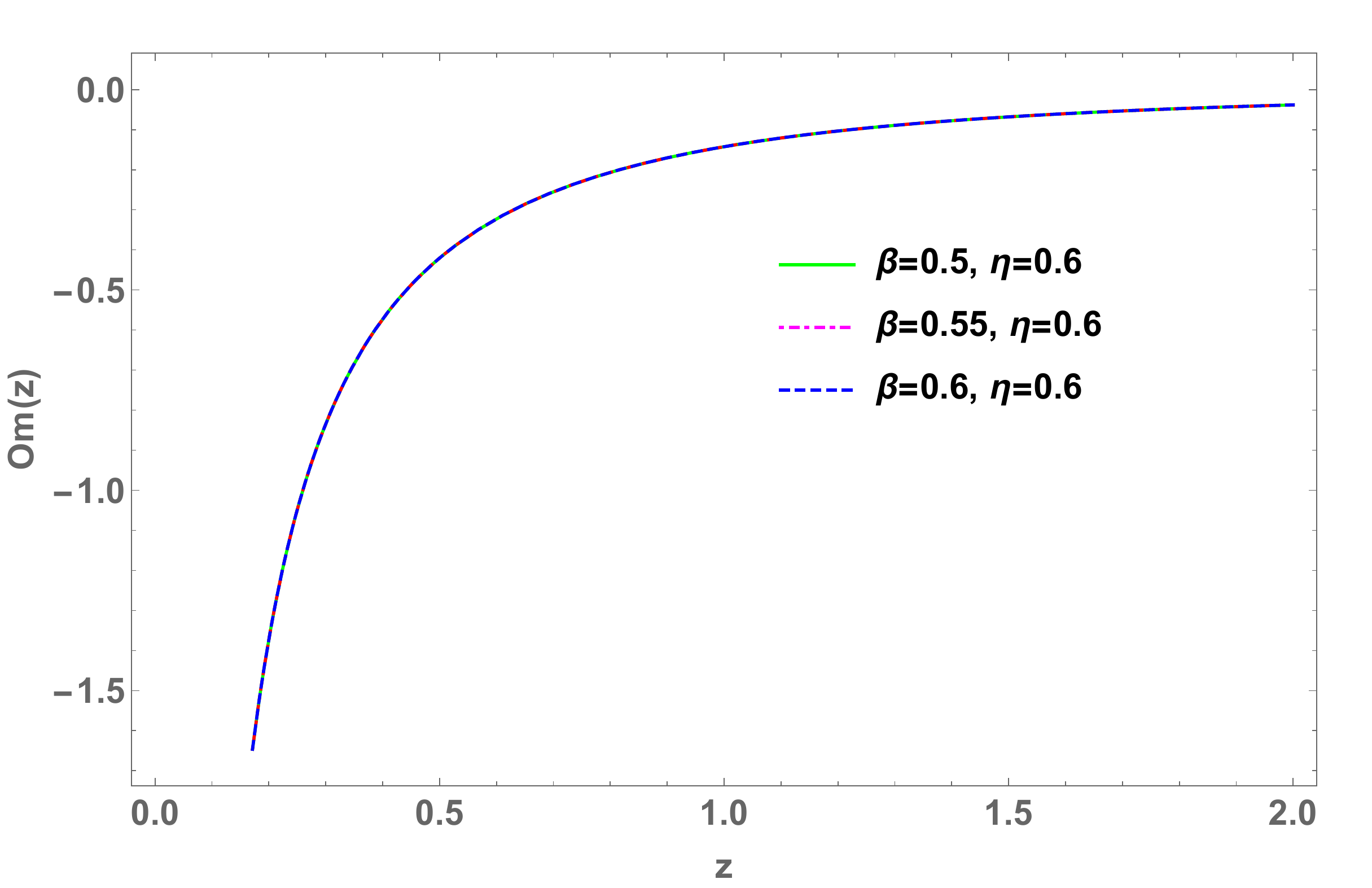}
  \caption{Variation of Om(z) against $z$ with $H_0=67.77$ km s$^{-1}$M pc$^{-1}$.}\label{fig5z}
\end{figure}

\section{Concluding Remarks}

In the presented manuscript we have discussed Friedmann-Lema\^itre-Robertson-Walker cosmological models in the context of the $f(R,T)$ gravity. 

Our cosmological solutions show a very healthy behaviour and yield great cosmological models. Particularly, let us argue about the EoS parameter evolution. Figs.6, 12 and 18 show a remarkable feature. They present for the evolution of $\omega$ a scenario which is consistent with three different stages of the universe evolution, namely radiation, matter and dark energy eras, as we argue below. 

One can see that for small values of time, $\omega\sim1/3$, which is the EoS parameter value for the primordial stage of the universe in which its dynamics was dominated by radiation \cite{ryden/2003}, whose high temperature did not allow, for a period of time, the formation of the first atoms. 

As the universe cool down, it allows the formation of the atoms and {\it a posteriori} the formation of stars, galaxies, clusters of galaxies etc. These objects, namely matter or pressureless matter, dominate the dynamics of the universe as a fluid with EoS $\omega=0$ \cite{ryden/2003}. From Figs.6, 12 and 18, we can see that after describing a radiation-dominated period, $\omega$ indeed passes through $0$, indicating the matter-dominated phase of the universe expansion. 

Finally, for high values of time, $\omega\rightarrow-1$, in accordance with recent observational data on fluctuations of temperature in the cosmic microwave radiation \cite{hinshaw/2013}. In standard model, the cosmological constant is the ``mechanism'' responsible for taking the universe to a ``dark energy''-dominated phase, in which a negative pressure fluid accelerates its expansion. In the present approach, rather, the extra terms in $f(R)$ and $f(T)$ are the responsible for such an important feature, which remarkably evades the cosmological constant problem \cite{weinberg/1989,peebles/2003,padmanabhan/2003}.

It is important to highlight that the description of three different stages of the evolution of the universe in a continuous and analytical form is not only a novelty in $f(R,T)$ gravity but also in the broad literature. Some of the few examples of complete cosmological models already present in the literature are those obtained from two scalar field quintessence models \cite{ms/2014} and decaying vacuum models \cite{lima/2013}.

By comparing our results with present literature we are led to conclude that our particular forms for $f(R)$ together with the linear term on $T$ are responsible for the remarkable features of the present model. On this regard, one can note that $f(R,T)$ functional forms which are linear on both $R$ and $T$-dependences generally do not yield complete cosmological scenarios as those here obtained \cite{pawar/2018,chaubey/2017,satish/2016}.

Moreover, In Fig.20, we plotted $Om(z)$ for the redshift range $0\leq z\leq 2$. We observe that when the redshift $z$ is increasing within the interval $0\leq z\leq 2$, the $Om(z)$ is monotonically increasing, which also indicates the accelerated expansion of the universe.

\begin{acknowledgements}
PKS and PS acknowledges DST, New Delhi, India for providing facilities through DST-FIST lab, Department of Mathematics, where a part of this work was done. The authors also thank the referee for the valuable suggestions, which improved the presentation of the present results.
\end{acknowledgements}

\end{document}